\documentclass{aa}

\usepackage{comment}
\usepackage{graphicx}
\usepackage[table,xcdraw,dvipsnames]{xcolor}
\usepackage{caption}
\usepackage{subcaption}

\usepackage{pdflscape} 
\usepackage{txfonts}

\usepackage{hyperref} 
\hypersetup{colorlinks=true, linktocpage=true, pdfstartpage=3,
            pdfstartview=FitV, breaklinks=true, pdfpagemode=UseNone,
            pageanchor=true, pdfpagemode=UseOutlines, plainpages=false,
            bookmarksnumbered, bookmarksopen=true, bookmarksopenlevel=1,
            hypertexnames=true, urlcolor=RoyalBlue,
            linkcolor=RoyalBlue, citecolor=RoyalBlue}

\usepackage[utf8]{inputenc}
\usepackage[T1]{fontenc}

\usepackage{har2nat}

\usepackage{booktabs} 
\usepackage{longtable}
\usepackage{supertabular}
\usepackage{amsmath}
\DeclareUnicodeCharacter{2212}{-}

\begin{document}

   \title{Star Formation and AGN Feedback in the Local Universe: Combining LOFAR and MaNGA}

   \author{C. R. Mulcahey
          \inst{1,2}\thanks{email: mulca23c@mtholyoke.edu}
          \and
          S. K. Leslie\inst{1,3}\thanks{email: leslie@strw.leidenuniv.nl }
          \and
          T. M. Jackson\inst{4}\inst{5}\inst{6}
          \and
          J. E. Young\inst{2}
          \and
          I. Prandoni\inst{7}
          \and
          M. J. Hardcastle\inst{8}
          \and
          N. Roy\inst{7}
          \and
          K. Ma\l{}ek \inst{10}\fnmsep\inst{11}
          \and
          M. Magliocchetti\inst{12}
          \and
          M. Bonato\inst{7}\fnmsep\inst{13}
          \and
          H. J. A. R\"{o}ttgering\inst{1}
          \and
          A. Drabent\inst{14}
          }

   \institute{ Leiden Observatory, Leiden University, PO Box 9513, NL-2300 RA Leiden, The Netherlands
        \and
            Mount Holyoke College, Department of Astronomy, 50 College St, South Hadley, MA 01075, USA
        \and 
            ARC Centre of Excellence for All Sky Astrophysics in 3 Dimensions (ASTRO 3D)
        \and
            Tuparev Astrotech, 3 Sofiyski Geroi Str, Sofia 1612, Bulgaria
        \and
            Astrosysteme Austria, Galgenau 19, 4212 Neumarkt im M{\"u}hlkreis, Austria
        \and
            Astronomisches Rechen-Institut, Zentrum f{\"u}r Astronomie der Universit{\"a}t Heidelberg, M{\"o}nchhofstr. 12-14, 69120 Heidelberg, Germany
        \and
            INAF ‑ Istituto di Radioastronomia and Italian ALMA Regional Centre, Via Gobetti 101, I-40129 Bologna, Italy
        \and
            Centre for Astrophysics Research, University of Hertfordshire, College Lane, Hatfield AL10 9AB
        \and
            Department of Astronomy $\&$ Astrophysics, University of California, 1156 High street, Santa Cruz, CA 95060
        \and
            National Centre for Nuclear Research, ul. Ho{\.z}a 69, 00-681, Warszawa, Poland
        \and 
            Aix Marseille Univ. CNRS, CNES, LAM, Marseille, France
        \and
            INAF-IAPS, Via Fosso del Cavaliere 100, 00133 Roma, Italy
        \and
            INAF ‑ Osservatorio Astronomico di Padova, Vicolo dell'Osservatorio 5, I-35122 Padova, Italy
        \and
            Th{\"u}ringer Landessternwarte (TLS), Sternwarte 5, 07778 Tautenburg, Germany
             }

   \date{}
 
  \abstract{
  The effect of Active Galactic Nuclei (AGN) on their host galaxies -- in particular their levels of star formation -- remains one of the key outstanding questions of galaxy evolution. Successful cosmological models of galaxy evolution require a fraction of energy released by an AGN to be redistributed into the interstellar medium to reproduce the observed stellar mass and luminosity function and to prevent the formation of over-massive galaxies. Observations have confirmed that the radio-AGN population is energetically capable of heating and redistributing gas at all phases, however, direct evidence of AGN enhancing or quenching star formation remains rare. With modern, deep radio surveys and large integral field spectroscopy (IFS) surveys, we can detect fainter synchrotron emission from AGN jets and accurately probe the star-forming properties of galaxies, respectively. In this paper, we combine data from the LOw Frequency ARray Two-meter Sky Survey with data from one of the largest optical IFS surveys, Mapping Nearby Galaxies at Apache Point Observatory to probe the star-forming properties of 307 local (z $<$ 0.15) galaxies that host radio-detected AGN (RDAGN). We compare our results to a robust control sample of non-active galaxies that each match the stellar mass, redshift, visual morphology, and inclination of a RDAGN host. We find that RDAGN and control galaxies have broad SFR distributions, typically lie below the star-forming main-sequence, and have negative stellar light-weighted age gradients. These results indicate that AGN selected based on their current activity are not responsible for suppressing their host galaxies' star formation. Rather, our results support the maintenance mode role that radio AGN are expected to have in the local Universe.
 }
            
\maketitle
%

\section{Introduction}
\label{section:intro}

How supermassive black holes (SMBHs) and their host galaxies coevolve has yet to be fully understood. During growth periods, in which SMBHs actively accrete gas and are known as active galactic nuclei (AGN), they can release an enormous amount of radiation across the entire electromagnetic spectrum and can form winds and jets in their host galaxies. Current cosmological models of galaxy evolution \citep[e.g. ][]{Bower2006, Schaye2015, Phillepich2019} require AGN to inject energy and momentum into their host galaxies' circumambient gas and interstellar medium (ISM) to reproduce the observed stellar mass and luminosity function and prevent the formation of over-massive galaxies. Observationally, the relation between star-formation (SF) history and the growth of SMBHs at the center of galaxies has been the subject of many studies \citep[e.g. ][and references therein]{Mullaney2012, Chen2013, Hickox2014} that have found that SF and black hole accretion rates (BHARs) are intimately tied at all redshifts \citep[e.g. ][]{Boyle1998, Aird2015}. This relationship likely indicates that SF and BH accretion share a common fuel source \citep[e.g. ][]{Silverman2009}. The correlation between the mass of the black hole and the stellar velocity dispersion \citep[M$_{BH} - \sigma_{*}$; e.g.][]{Haehnelt2000} as well as the link between  M$_{BH}$ and the mass of the stellar bulge \citep[M$_{BH} - M_{buldge}$; e.g.][]{Harring2004}, further hint at the co-evolution of black holes and stellar bulges, thereby suggesting a link between BHARs and star-formation rates (SFRs). However, studies investigating the relation between AGN activity and star-forming activity \citep[e.g.][]{Netzer2009, Rosario2012, Gurkan2015, Stanley2015, Jackson2020} have so far yielded mixed results.  

There are two prominent ways that AGN feedback can affect its host galaxy. Outflows from AGN can enhance SF (positive feedback) by compressing molecular clouds \citep[e.g. ][]{Schaye2015} and/or the interstellar medium \citep[e.g, ][]{Ishibashi2012}. Direct evidence of positive feedback is rare \citep[e.g. ][]{Cresci2015a, Shin2019, Nesvadba2020} and is typically observed in a companion satellite along the host galaxy's radio axis \citep[e.g. ][]{Klamer2004, Croft2006, Feain2007, RodriguezZaurin2007, Elbaz2009, Crockett2012, Gilli2019}. Conversely, AGN can suppress SF (negative feedback) via mechanical energy from winds, outflows, or jets heating the surrounding ISM and preventing molecular gas from radiatively cooling or due to AGN-driven outflows expelling gas from the host galaxy \citep[e.g.][]{Binney1995, Ciotti2001, Croton2006, Ciotti2007, McNamara2007, Nesvadba2008, Cattaneo2009, Ciotti2010,  Nesvadba2010, Fabian2012, Yuan2014, Heckman2014}. On longer timescales, jets can heat the circumgalactic and halo gas, preventing the cooling of gas and future SF \citep[e.g.][]{Ciotti2001, Ciotti2007, McNamara2007}. Furthermore, the role of AGN feedback varies depending on the type of AGN the galaxy hosts. Radio-loud AGN can either be radiatively efficient or radiatively inefficient. Radiatively efficient AGN are typically connected to the most luminous AGN and accrete gas close to the Eddington limit from an optically thick, geometrically thin accretion disk. Radio-loud Quasi-Stellar Objects (QSOs) and high excitation radio galaxies (HERGs) -- further classifications of radiatively-efficient AGN -- are capable of producing powerful, two-sided jets that produce synchrotron radiation detectable at radio wavelengths. Energy released from the accretion disk may be capable of driving massive outflows gas and ultimately remove it from the potential well \citep[e.g. ][]{Cattaneo2009, Fabian2012}. Conversely, radiatively inefficient AGN -- also referred to as low excitation radio galaxies (LERGs) -- are linked to low to intermediate luminosity AGN and contain a geometrically-thick, advection-dominated accretion flows, which can also produce powerful radio jets. Radiatively-inefficient AGN have been shown to inject heat into their surroundings at a rate that is commensurate with the rate of cooling from
the intergalactic medium, and are responsible for maintaining galaxy quiescence \citep[e.g.][]{Binney1995, Ciotti2001, Bower2006, Ciotti2007, McNamara2007, Cattaneo2009, Ciotti2010, Fabian2012, Yuan2014, Heckman2014, Smolvcic2017, Hardcastle2019}.

Significant advances in our understanding of the effect of radio-mode AGN on their host galaxies have been achieved by coupling radio surveys such as the National Radio Astronomy Observatory (NRAO) Very Large Array Sky Survey \citep[NVSS; 1.4 GHz continuum; ][]{Condon1998}, the Faint  Images of the Radio Sky at Twenty centimeters \citep[FIRST; 1.4 GHz continuum; ][]{Becker1995}, the Very Large Array Sky Survey \citep[VLASS; 2-4 GHz; ][]{Hales2013}, and Tata Institute of Fundamental Research (TIFR) Giant Metrewave Radio Telescope (GMRT) Sky Survey \citep[TGSS; 150 MHz; ][]{Intema2017} with optical spectroscopic surveys such as the Sloan Digital Sky Survey \citep[SDSS; ][and references therein]{York2000, Stoughton2002} and the Two-degree-Field Galaxy Redshift Survey \citep[2dFGRS; ][]{Colless2001}. Statistical studies that have combined these surveys \citep[e.g. ][]{Best2005a, Best2005b, Sadler2002} have improved our understanding of the physical properties and prevalence of radio-AGN activity, but the nature of AGN emitting at radio frequencies lower than 1.4 GHz are yet to be fully understood. \citet{Sabater2019} combined data from the first data release (DR1) of the Low-Frequency Array \citep[LOFAR; 10-240 MHz; ][]{vanHaarlem2013} Two-Metre Sky Survey \citep[LoTSS; ][]{Shimwell2017} with optical spectroscopic data from SDSS DR7 \citep{Abazajian2009} and found that the most massive AGN host galaxies ($>$10$^{11}$M$_{*}$) always exhibit radio-AGN activity. These results suggest that radio-AGN activity is dictated by the host galaxy's fuel supply and that radio-AGN play a significant role in maintaining quiescence.   
 
Simultaneously, integral field spectroscopy (IFS) surveys are revolutionizing our understanding of AGN by enabling more detailed investigations than previously possible. Unlike long-slit spectroscopy, which obtains a spectrum for a single point in the galaxy or acquiring spectra along a ``slice" of the galaxy, IFS obtains resolved, two-dimensional spectra across the surface of the galaxy. IFS, in combination with stellar population modelling, permits the spatially resolved study of a galaxy’s properties such as current SFRs, metallicities, and stellar ages. Moreover, the gas and stellar kinematics over an entire galaxy can be obtained, enabling the effect of winds and dynamical disturbances to be examined. One of the largest, optical IFS survey is the Mapping Nearby Galaxies at Apache Point Observatory \citep[MaNGA; ][]{Bundy2015} survey, which is one of three core parts of the fourth phase of SDSS (SDSS-IV). MaNGA has acquired observations with a spatial resolution of 2$^{\prime\prime}$.5 for $\sim$ 10,000 unique, low-redshift (0.01 $<$ \textit{z} $<$ 0.15; median \textit{z} = 0.03), massive (M$_{*}$ $>$ 10$^{9}$ M$_{\odot}$) galaxies \citep{Yan2016b}.  Previous MaNGA AGN studies underscore the importance of spatially resolved measurements to provide unprecedented insight on the prevalence and properties of AGN and their host galaxies \citep[e.g. ][]{Rembold2017, Sanchez2018, Wylezalek2018, Comerford2020, Wylezalek2020}. Moreover, multiple IFS studies \citep[e.g.][]{Sanchez2018, Comerford2020, Wylezalek2020, Venturi2021} have found evidence for AGN driving outflows, turbulence, and suppressing SF over time.

In this study, we build and improve on these previous works by coupling IFS data from MaNGA DR16 \citep[][]{Bundy2015, Ahumada2020} with data from the second data release of LoTSS (Shimwell et al in prep.). By leveraging the unique capabilities of LOFAR, our sample contains fainter radio-AGN -- as well as remnant emission from sources that have recently shut off their jet activity -- than those that have been previously analyzed with MaNGA data. We will determine where in relation to the SFMS the AGN host galaxies and non-active galaxies lie, compare the distribution of SFRs in regions ionized by hot stars, and will investigate how the age of stellar populations in AGN galaxies and non-active galaxies change as a function of galactocentric radius. We describe the sample and data used to achieve our research goals in Section \ref{section:Sample and data}. After outlining the methods used to define the radio-detected AGN (RDAGN) and control sample in Section \ref{section:Sample seelction}, we determine these galaxies' relation to the SF main-sequence (SFMS) in Section \ref{section:Relation to the SFMS}. In Sections \ref{section:Spatially resolved stellar and nebular gas properties} and \ref{section:Stellar age gradient}, we examine the spatially resolved properties of the stellar and nebular gas populations and probe their stellar light-weighted age gradients, respectively. Finally, we discuss our interpretation of these results and present a summary of our conclusions in Sections \ref{section:Discussion} and \ref{section:Conlcusions}. Throughout this work, we assume the cosmological parameters of H$_{0} =$ 70 km s$^{-1}$, $\Omega_{M} =$ 0.3 and a Salpeter initial mass function \citep[IMF; ][]{Salpeter1955}.

 \begin{figure*}[!t]
    \centering
    \includegraphics[width = \textwidth, trim= 1cm 2cm 1cm 1cm, clip]{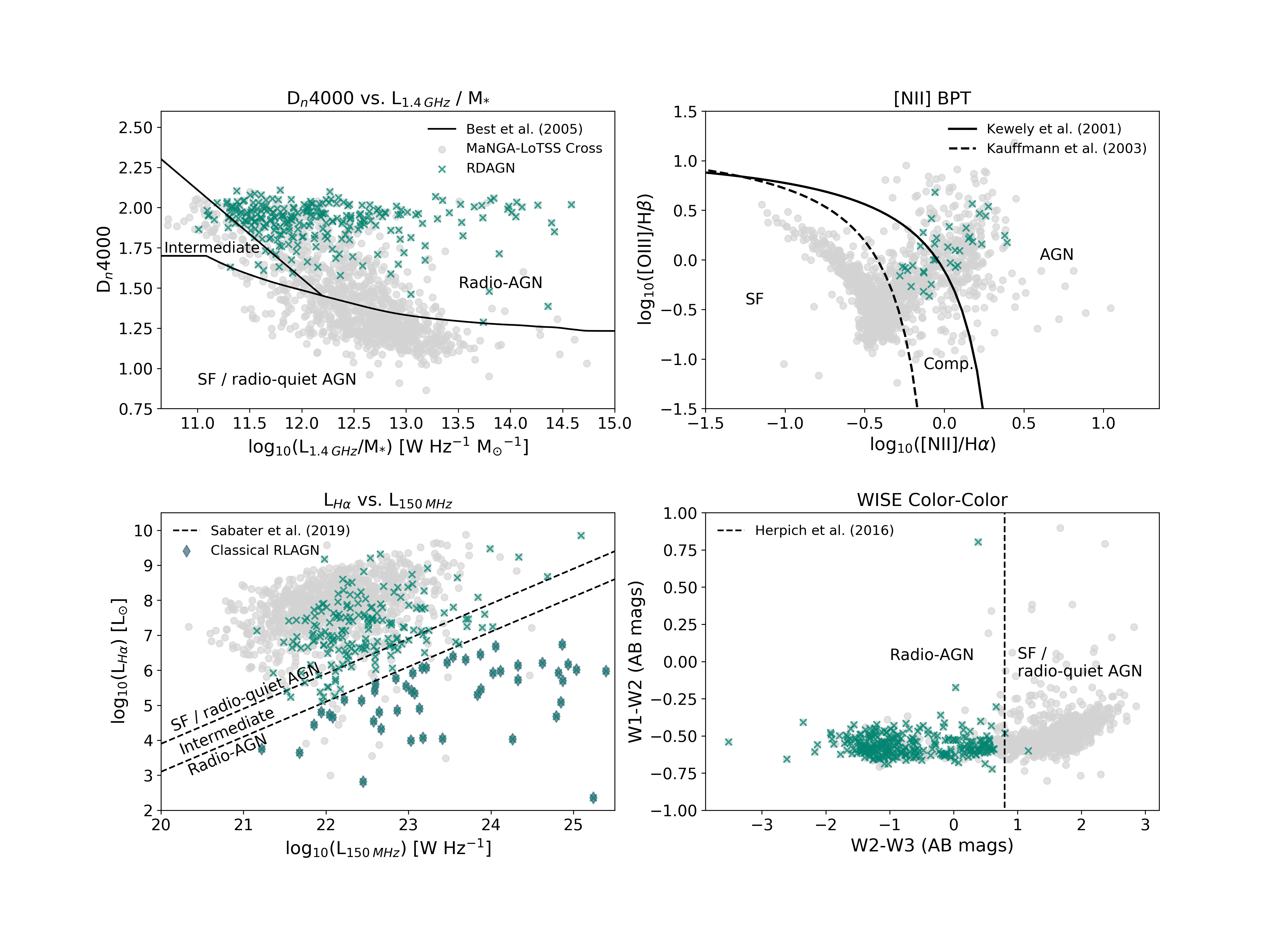}
    \caption{Location of the RDAGN host galaxies on the four diagnostic diagrams used to separate galaxies whose radio emission was from SF from those galaxies likely powered by AGN.  \textit{Top row from left to right: }D$_{n}$4000 vs. L$_{1.4\;\mathrm{GHz}}$/M$_{*}$ from \citet{Best2005b}, the [NII]/H$\alpha$ BPT diagram \citep{Baldwin1981}. \textit{Bottom row from left to right: }L$_{\mathrm{H\alpha}}$ vs. L$_{150\;\mathrm{MHz}}$, WISE W1-W2 vs. W2-W3 color-color diagram. The lines in each diagram represent division between SF/radio-quiet AGN, intermediate, and radio AGN. The final RDAGN sample, obtained following our criteria described in Section \ref{section:Sample and data}, is indicated by green ``x's." The grey circles represent the full sample of MaNGA-LoTSS galaxies. Classical RLAGN are represented on the L$_{\mathrm{H\alpha}}$ vs. L$_{150\;\mathrm{MHz}}$ diagram with dark grey diamonds.}
    \label{fig:RDAGN Diagnotstic Diagrams}
\end{figure*}

\section{Sample and data}
\label{section:Sample and data}
This work makes use of the second data release of LoTSS, which is an ongoing radio continuum (120-168 MHz) survey of the northern sky. The scientific objectives of LoTSS are to exploit the unique capabilities of LOFAR to shed new light on the formation and evolution of massive black holes, cluster galaxies, and the high redshift Universe \cite[see][and references therein]{Shimwell2019}. LoTSS uses LOFAR's high band antennas (HBA) and aims to reach a sensitivity $<$ 0.1 mJy beam$^{-1}$ at an angular resolution of $\sim$ 6$^{\prime\prime}$. LoTSS DR2 (Shimwell et al. in prep.) covers 27\% of the northern sky and is composed of two discrete fields -- denoted the 0h and 13h fields -- covering 5700 deg$^{2}$ in total (1480 deg$^{2}$ in the 0h field and 4240 deg$^{2}$ in the 13h field). The astrometric accuracy of the images is $\sim$ 0.2$^{\prime\prime}$. The flux calibration of LOFAR DR2 is uncertain to $<$10$\%$. 

MaNGA uses the Baryon Oscillation Spectroscopic Survey (BOSS) spectrograph \citep{Smee2013} on the 2.5-meter telescope at Apache Point Observatory \citep[APO; ][]{Gunn2006} to obtain high-resolution (R $\sim$ 2000) spectra over a large wavelength range (3600 - 10300 \text{\r{A}}). MaNGA uses integral field units (IFUs) that consist of tightly-packed hexagonally bundled 2$^{\prime\prime}$ fibers, which have five different sizes (19, 37, 61, 91 and 127 fibers) corresponding to physical diameters of 12$^{\prime\prime}$, 17$^{\prime\prime}$, 22$^{\prime\prime}$, 27$^{\prime\prime}$, and 32$^{\prime\prime}$ \citep{Drory2015}. Raw fiber spectra have a calibration accuracy better than 5$\%$ \citep{Yan2016a, Yan2016b}. We use MaNGA observations from the sixteenth data release of SDSS-IV, which includes observations of 4824 galaxies taken before August 2018 \citep[][]{Ahumada2020}. Final data cubes and row stacked spectra (RSS) were produced using the MaNGA DRP \citep[][]{Law2016}. Global emission line fluxes used in this study were obtained from the Portsmouth Group \citep{Thomas2013}. In addition, we use measured galaxy properties from several MaNGA Value Added Catalogs (VACs). We relied on the Pipe3D VAC \citep{Sanchez2016a, Sanchez2016b, Sanchez2018} for cumulative stellar mass (M$_{*}$) measurements, SFR (obtained from stellar population modeling), and stellar, light-weighted age gradient ($\alpha$; slope of the gradient of the luminosity-weighted log-age of the stellar population within a galactocentric distance of 0.5 to 2.0 R$_{e}$). To determine the morphological classifications of galaxies, we use T-TYPE values from the Morphology Deep Learning DR15 Value Added Catalog \citep[VAC; ][]{DomSanchez2018}. We probe the environment in which these galaxies reside using measurements from the Galaxy Environment for MaNGA (GEMA) VAC \citep[Argudo-Fernádez et al. in prep.]{Argudo-Fernandez2015}. Finally, D$_{n}$4000 and M$_{*}$ values used in the D$_{n}$4000 vs. L$_{1.4\;\mathrm{GHz}}$/M$_{*}$ diagram (see Section \ref{subsec:Selecting radio-detected AGN}) were taken from the MPA-JHU VAC\footnote{VAC created by the Max Planck for Astrophysics (MPA) and Johns Hopkins University (JHU) groups.}.  

Wide-Field Infrared Survey Explorer \citep[WISE; ] []{Wright2010} data used in this work come from unWISE forced photometry performed by \citet{Lang2016} on un-blurred co-added WISE images \citep{Lang2014b} at over 400 million optical SDSS source positions. Therefore, the unWISE data are naturally connected to the SDSS parent sample from which MaNGA targets were drawn.

The sky position of the LoTSS and MaNGA catalogs were matched using Tool for OPerations on Catalogues And Tables (TOPCAT). We matched the two source catalogs using a 5$^{\prime\prime}$ matching radius in RA and Dec. We estimate the total fraction of spurious matches to be $<$10\% based on the average of 15 simulated MaNGA catalogs with randomized positions. Based on the cross-matching criteria, there are 1410 sources detected between the LoTSS and MaNGA survey. We use the SDSS spectroscopic ID (specObjID) to cross match the MaNGA-LoTSS catalog with the MPA-JHU catalog and with the global emission line flux catalog. For the Pipe3D, Morphology Deep Learning, and GEMA value added catalogs, we used the MANGAID identifier for cross matching. 

\section{Sample selection and properties}
\label{section:Sample seelction}

\subsection{Selecting radio-detected AGN}
\label{subsec:Selecting radio-detected AGN}

\begin{table}
\centering
\resizebox{\columnwidth}{!}{%
\begin{tabular}{@{}cccccc@{}}
\toprule
\begin{tabular}[c]{@{}c@{}}D$_{n}$4000 vs. \\ L$_{1.4\;\mathrm{GHz}}$ / M$_{*}$\end{tabular} &
  [NII] BPT &
  \begin{tabular}[c]{@{}c@{}}L$_{\mathrm{H\alpha}}$ vs.\\ L$_{150\;\mathrm{MHz}}$\end{tabular} &
  \begin{tabular}[c]{@{}c@{}}WISE\\ Color-Color\end{tabular} &
  \begin{tabular}[c]{@{}c@{}}Number\\ of Galaxies\end{tabular} &
  \begin{tabular}[c]{@{}c@{}}Final\\ Classification\end{tabular} \\ \midrule
SF  & SF  & SF  & \multicolumn{1}{c|}{SF}  & \multicolumn{1}{c|}{443} & SF  \\
AGN & Unc & SF  & \multicolumn{1}{c|}{AGN} & \multicolumn{1}{c|}{142} & AGN \\
AGN & Int & SF  & \multicolumn{1}{c|}{SF}  & \multicolumn{1}{c|}{106} & Unc \\
AGN & SF  & SF  & \multicolumn{1}{c|}{SF}  & \multicolumn{1}{c|}{87}  & SF  \\
Unc & Unc & Unc & \multicolumn{1}{c|}{SF}  & \multicolumn{1}{c|}{84}  & SF  \\
AGN & Unc & SF  & \multicolumn{1}{c|}{SF}  & \multicolumn{1}{c|}{79}  & SF  \\
Unc & Unc & Unc & \multicolumn{1}{c|}{AGN} & \multicolumn{1}{c|}{61}  & AGN \\
AGN & Unc & AGN & \multicolumn{1}{c|}{AGN} & \multicolumn{1}{c|}{60}  & AGN \\
Int & Unc & SF  & \multicolumn{1}{c|}{AGN} & \multicolumn{1}{c|}{42}  & Unc \\
SF  & Int & SF  & \multicolumn{1}{c|}{SF}  & \multicolumn{1}{c|}{39}  & SF  \\
AGN & AGN & SF  & \multicolumn{1}{c|}{SF}  & \multicolumn{1}{c|}{37}  & Unc \\
AGN & Unc & Int & \multicolumn{1}{c|}{AGN} & \multicolumn{1}{c|}{32}  & AGN \\
AGN & AGN & SF  & \multicolumn{1}{c|}{AGN} & \multicolumn{1}{c|}{23}  & AGN \\
AGN & Unc & Unc & \multicolumn{1}{c|}{AGN} & \multicolumn{1}{c|}{23}  & AGN \\
Unc & Unc & Unc & \multicolumn{1}{c|}{Unc} & \multicolumn{1}{c|}{21}  & Unc \\
AGN & Int & SF  & \multicolumn{1}{c|}{AGN} & \multicolumn{1}{c|}{13}  & AGN \\
SF  & SF  & SF  & \multicolumn{1}{c|}{Unc} & \multicolumn{1}{c|}{10}  & SF  \\
Int & Int & SF  & \multicolumn{1}{c|}{SF}  & \multicolumn{1}{c|}{10}  & Unc \\
SF  & Unc & SF  & \multicolumn{1}{c|}{SF}  & \multicolumn{1}{c|}{10}  & SF  \\
AGN & Unc & Unc & \multicolumn{1}{c|}{SF}  & \multicolumn{1}{c|}{8}   & Unc \\
SF  & AGN & SF  & \multicolumn{1}{c|}{SF}  & \multicolumn{1}{c|}{7}   & SF  \\
AGN & Unc & Int & \multicolumn{1}{c|}{SF}  & \multicolumn{1}{c|}{7}   & Unc \\
Int & SF  & SF  & \multicolumn{1}{c|}{SF}  & \multicolumn{1}{c|}{7}   & SF  \\
SF  & Unc & Unc & \multicolumn{1}{c|}{SF}  & \multicolumn{1}{c|}{6}   & SF  \\
Int & Int & SF  & \multicolumn{1}{c|}{AGN} & \multicolumn{1}{c|}{5}   & AGN \\
Int & AGN & SF  & \multicolumn{1}{c|}{SF}  & \multicolumn{1}{c|}{5}   & Unc \\ \midrule
\multicolumn{4}{c}{Total}                  & 380                      & AGN \\
\multicolumn{4}{c}{Total}                  & 783                      & SF  \\
\multicolumn{4}{c}{Total}                  & 247                      & Unc \\ \bottomrule
\end{tabular}%
}
\caption[Classification of galaxies detected in the MaNGA and LoTSS surveys.]{Number of galaxies and their overall classification for different combinations of the four diagnostic methods of Figure \ref{fig:RDAGN Diagnotstic Diagrams}. We show only the combinations that classified at least five galaxies to save space. In each diagnostic diagram, galaxies whose emission is dominated by AGN activity or SF are classified as ``AGN” and ``SF”, respectively. ``Int”  indicates both AGN activity and SF contribute to galaxy’s emission, and ``Unc” means that there were no measurements for those galaxies to be classified.}
\label{tab: Galaxy classification}
\end{table}

\begin{table}
\centering
\resizebox{\columnwidth}{!}{%
\begin{tabular}{@{}rcccc@{}}
\toprule
\multicolumn{1}{c}{{\color[HTML]{333333} Diagnostic Method}} & \multicolumn{4}{c}{{\color[HTML]{333333} \begin{tabular}[c]{@{}c@{}}Number Classified in MaNGA-LoTSS Cross\\\color[HTML]{018571} (Number in RDAGN Sample)\end{tabular}}} \\ \midrule
\multicolumn{1}{c}{{\color[HTML]{333333} }} & {\color[HTML]{333333} AGN} & {\color[HTML]{333333} SF} & {\color[HTML]{333333} Intermediate} & {\color[HTML]{333333} Unclassified} \\
{\color[HTML]{333333} D$_{n}$4000 vs. L$_{1.4\;\mathrm{GHz}}$ / M$_{*}$} & {\color[HTML]{333333} 633} & {\color[HTML]{333333} 525} & {\color[HTML]{333333} 86} & {\color[HTML]{333333} 166} \\
 & {\color[HTML]{018571} (247)} & {\color[HTML]{018571} (0)} & {\color[HTML]{018571} (11)} & {\color[HTML]{018571} 49} \\
{\color[HTML]{333333} [NII] BPT} & {\color[HTML]{333333} 77} & {\color[HTML]{333333} 555} & {\color[HTML]{333333} 177} & {\color[HTML]{333333} 601} \\
 & {\color[HTML]{018571} (23)} & {\color[HTML]{018571} (0)} & {\color[HTML]{018571} (14)} & {\color[HTML]{018571} (307)} \\
{\color[HTML]{333333} L$_{\mathrm{H\alpha}}$ vs. L$_{150\;\mathrm{MHz}}$} & {\color[HTML]{333333} 68} & {\color[HTML]{333333} 1085} & {\color[HTML]{333333} 46} & {\color[HTML]{333333} 211} \\
 & {\color[HTML]{018571} (52)} & {\color[HTML]{018571} (149)} & {\color[HTML]{018571} (33)} & {\color[HTML]{018571} (73)} \\
{\color[HTML]{333333} WISE Color-Color} & {\color[HTML]{333333} 426} & {\color[HTML]{333333} 942} & {\color[HTML]{333333} $-$} & {\color[HTML]{333333} 42} \\
 & {\color[HTML]{018571} (302)} & {\color[HTML]{018571} (2)} & {\color[HTML]{018571} $-$} & {\color[HTML]{018571} (3)} \\ \bottomrule
\end{tabular}
}
\caption[Classification of galaxies.]{Number of galaxies classified for each category (AGN, SF, Intermediate, Unclassified) for each diagnostic diagram. The green, bracketed numbers on the second row for each diagnostic represent the number of galaxies with an overall classification as RDAGN.}

\label{tab:Galaxy classification}
\end{table}

The most reliable method for building a sample of pure radio-AGN is to select objects whose radio luminosity greatly surpasses that from their SF \citep{Hardcastle2016, CalistroRivera2017, Smolvcic2017, Hardcastle2019}. In this work, we chose to separate our AGN host galaxies from SF galaxies using global optical emission-line properties, radio luminosities, and mid-infrared luminosities following the approach used by \citet{Sabater2019}. We chose to take a multi-wavelength approach in order to build a complete AGN sample with varying host galaxy properties. However, we highlight a sub-sample of classical radio-loud AGN (RLAGN), which is composed of RDAGN whose radio emission is higher than what is expected from their SFR alone based on our third diagnostic technique.

The first technique is the D$_{n}$4000 vs. L$_{1.4\;\mathrm{GHz}}$/M$_{*}$ diagram, which was developed by \citet{Best2005a}, and it is shown in the upper upper left panel of Figure \ref{fig:RDAGN Diagnotstic Diagrams}. This method uses the strength of the 4000 \text{\r{A}} break (D$_{n}$4000) in each galaxy's spectrum as a function of the ratio of the radio luminosity to stellar mass. This diagnostic diagram was developed using 1.4 GHz data, so we converted the LoTSS radio luminosity from 150 MHz to 1.4 GHz by assuming the established spectral index value of $\alpha =$ 0.7 \citep[S$_{v}$ $ \propto$ $\nu^{-\alpha}$; ][]{Condon2002, Smolvcic2017}. We use D$_{n}$4000 and M$_{*}$ values from the MPA-JHU VAC. \citet{Best2005a} demonstrated that because D$_{n}$4000 and L$_{1.4\;\mathrm{GHz}}$/M$_{*}$ both depend on the SFR of galaxies, SF galaxies will populate a similar region in the D$_{n}$4000 vs. L$_{1.4\;\mathrm{GHz}}$/M$_{*}$ plane. Moreover, SF galaxies can be separated from AGN host galaxies because they will typically have a weaker D$_{n}$4000 values than AGN host galaxies of a comparable radio luminosity. The curved division line between SF / radio-quiet AGN and radio-AGN represents the 3 Gyr exponential SF track \citep{Best2005a}.\footnote{The tracks used in this study were provided by Philip Best.} At D$_{n}$4000 $>$ 1.7 we replace the 3 Gyr exponential star formation track with a horizontal line, as proposed by \citet{Sabater2019}. The purpose of the addition is to avoid misclassifying AGN galaxies with large D$_{n}$4000 values as SF galaxies. The second diagnostic line is defined by \(D_{n}4000 = 1.45-0.55\times(L_{1.4\;GHz}/M_{*}-12.2)\) \citep[]{Best2012}. All sources that lie above the 3 Gyr exponential SF track and to the right of this second line are classified as radio-AGN. Conversely, galaxies that fall above the 3 Gyr exponential SF track and to the left of the second diagnostic line are intermediate, which means that both SF and AGN activity likely contribute to the radio emission. Finally, all sources that lie below the 3 Gyr exponential SF track are classified as SF / radio-quiet AGN.

The second technique that we use to separate AGN galaxies from SF galaxies is [NII] Baldwin, Phillips $\&$ Telervich \citep[BPT; ][]{Baldwin1981} diagram, which is shown in the upper right panel of Figure \ref{fig:RDAGN Diagnotstic Diagrams}. For this diagnostic, we use global emission line fluxes that were obtained by the Portsmouth Group \citep[][]{Thomas2013}. The diagram utilizes the ratio of narrow lines [OIII] $\lambda$5007 to H$\beta$ and [NII] $\lambda$6583 to H$\alpha$ to separate SF galaxies, from composite galaxies (a mix of ionizing sources likely contribute to the emission), from AGN galaxies. These line ratios can separate SF galaxies from AGN galaxies because the emission lines are affected by the hardness of the ionizing radiation field and the ionizing parameter. AGN galaxies will therefore have enhanced [NII]/H$\alpha$ ratios because they have a harder ionizing radiation field than SF galaxies. The first diagnostic line on this diagram, represented by the solid, black line in Figure \ref{fig:RDAGN Diagnotstic Diagrams}, is the maximum starburst line from \citet{Kewley2001}, which is defined by $(\log(\mathrm{[OIII]/H}\beta) < 0.61/(\log(\mathrm{[NII]/H}\alpha) − 0.47) + 1.19)$. Unlike \citet{Sabater2019}, we include the ``composite" classification on the [NII] BPT (classification "Int" in Table \ref{tab: Galaxy classification}). This second diagnostic line, represented by the dashed, black line in Figure \ref{fig:RDAGN Diagnotstic Diagrams} separates pure SF galaxies from composite galaxies \citep{Kauffmann2003} and is defined by $(\log(\mathrm{[OIII]/H}\beta) < 0.61/(\log(\mathrm{[NII]/H}\alpha) − 0.05) + 1.3\). Using the Kauffman line results in a more complete AGN-host selection than the Kewley classification, but it is far from a pure AGN selection as, for example, hot low-mass evolved stars, and shock ionization can also produce composite line ratios \citep[e.g. ][]{Sanchez2020}. For our [NII] BPT classification, we also require all galaxies to have EW(H$\alpha$)$>$3 \text{\r{A}} to avoid passive galaxies whose ionization is dominated by old stars (\citealt[e.g. ][]{Stasinska2008}).

A limitation of the global BPT diagrams is that average or integrated emission line ratios are affected by various galactic properties; galaxies are rarely only ``star-forming'' or ``AGN'' or ``quiescent''. Extinction may bias this selection, but because emission-line ratios are close together in wavelength ([NII] and H$\alpha$) and ([OIII] and H$\beta$), we expect similar extinction values for each line and therefore do not expect extinction to significantly bias our results. The optical narrow-line ratios of Type 1 AGN will have lower [NII]/H$\alpha$ values than Type 2 AGN because the AGN are unobscured and the narrow emission lines are ``blended" with broad emission lines \citep[e.g. ][]{Zhang2008, Stern2013}. There is only one Type 1 AGN in our final RDAGN sample (plateifu 8549-12702), which we identified using the SDSS-DR7 Type 1 AGN catalog developed by \citet{Oh2015}. Emission line ratios can also be enhanced by other non-AGN activity, such as Wolf-Rayet stars \citep[e.g. ][]{Brinchmann2008}, post-asymptotic giant branch stars \citep[e.g.][]{Binette1994, Yan2012, Belfiore2016}, and shocks driven by galaxy mergers, jets, and stellar winds \citep[e.g.][]{Rich2011,Kewley2013}. We explore these other mechanisms in Section \ref{subsec:Excitation maps}.

Our third technique, which is shown in the bottom left panel of Figure \ref{fig:RDAGN Diagnotstic Diagrams}, is the relation between the luminosity of H$\alpha$ (L$_{\mathrm{H\alpha}}$) and the LoTSS radio luminosity (L$_{150\;\mathrm{MHz}}$). Using the global emission line fluxes measured by the Portsmouth Group \citep{Thomas2013}, we measured the dust-corrected L$_{\mathrm{H\alpha}}$ using the average, R$_{V}$-dependent extinction function from \citet{Cardelli1989} and assume R$_{V}$ to be 3.1 \citep[][]{Savage1979, Cardelli1989}. Direct measurements of a galaxy's SFR can be determined from L$_{\mathrm{H\alpha}}$ and, in the absence of an AGN, L$_{150\;\mathrm{MHz}}$. Therefore, the locus of SF galaxies on the L$_{\mathrm{H\alpha}}$ vs. L$_{150\;\mathrm{MHz}}$ diagram is separate from the locus of AGN host galaxies. The diagnostic lines to separate SF galaxies from AGN galaxies are adopted from \citet{Sabater2019}: \(\log_{10}(L_{H\alpha}/L_{\odot}) = \log_{10}(L_{150\;\mathrm{MHz}}/\mathrm{W Hz}^{{-}1}) {-} 16.9\) and \(\log_{10}(L_{H\alpha}/L_{\odot}) = \log_{10}(L_{150\;\mathrm{MHz}}/\mathrm{W Hz}^{{-}1}) {-} 16.1\). Galaxies that lie below the bottom diagnostic line are classified as radio-AGN, intermediate if the galaxies lie between the two lines, and SF / radio-quiet AGN if the galaxies lie above the top diagnostic line. Our classical RLAGN sub-sample (52 galaxies in total) consists of the RDAGN host galaxies classified as a radio-AGN on this diagram (represented by the grey diamonds).

Our final method of separating AGN host galaxies from SF galaxies is the W1-W2 vs. W2-W3 mid-infrared WISE colors diagnostic diagram, which is shown in the lower right panel of Figure \ref{fig:RDAGN Diagnotstic Diagrams}. We obtain the mid-infrared WISE colors from the unWISE forced photometry catalog of 400 million SDSS sources \citet{Lang2016}. WISE colors are useful for detecting both obscured and unobscured AGN because hot dust surrounding AGN radiates strongly in mid-infrared emission. Following \citet{Sabater2019}, we use the division from \citet{Herpich2016} where galaxies with W2-W3$<$0.8 mag (AB) are radio-AGN.

\begin{figure}[!t]
     \centering
     \begin{subfigure}[b]{0.46\textwidth}
         \centering
         \includegraphics[width=\textwidth, trim=0 0cm 0 0 cm]{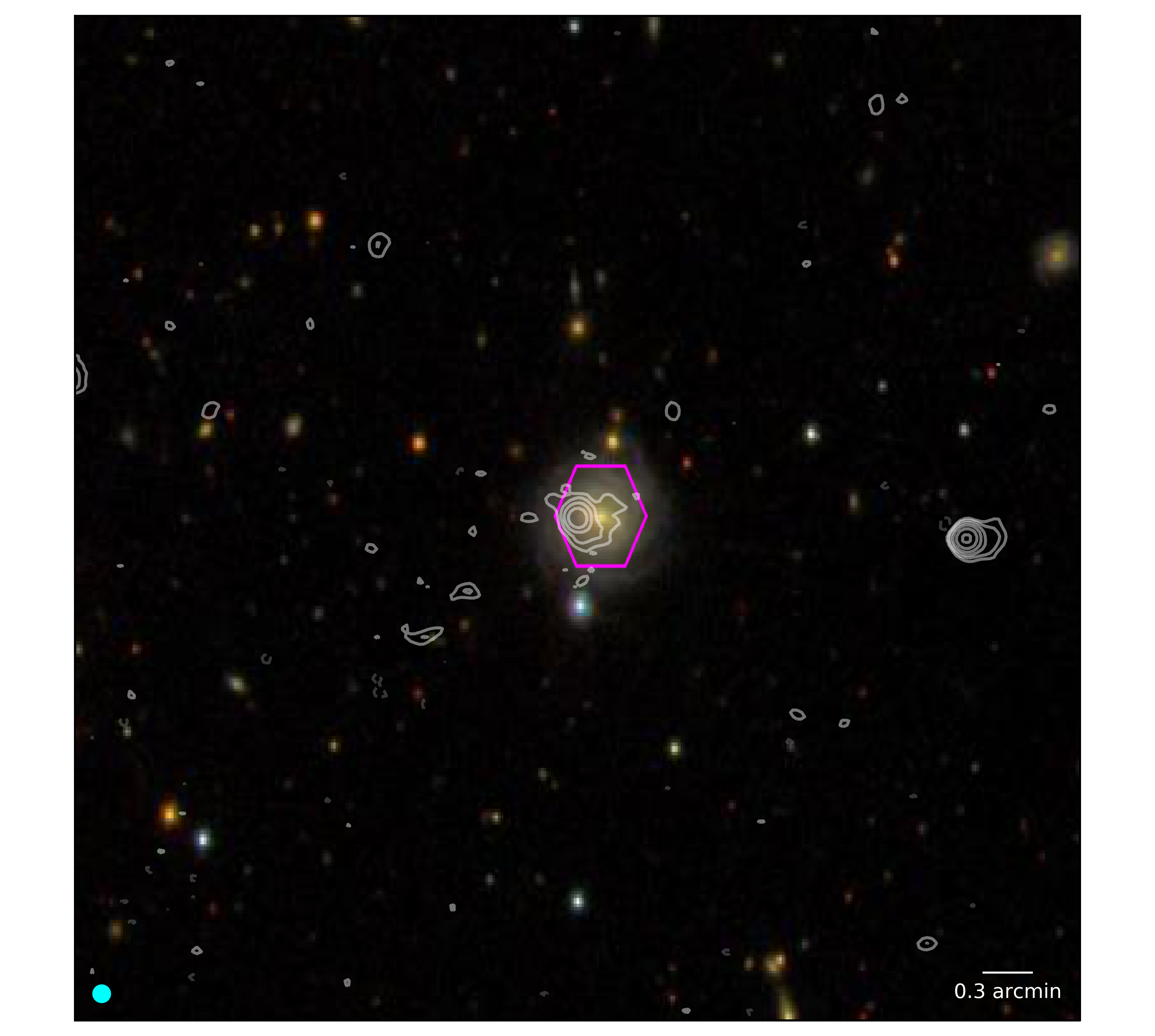}
     \end{subfigure}
     \vfill
     \begin{subfigure}[b]{0.46\textwidth}
         \centering
         \includegraphics[width=\textwidth, trim=0 0cm 0 0cm]{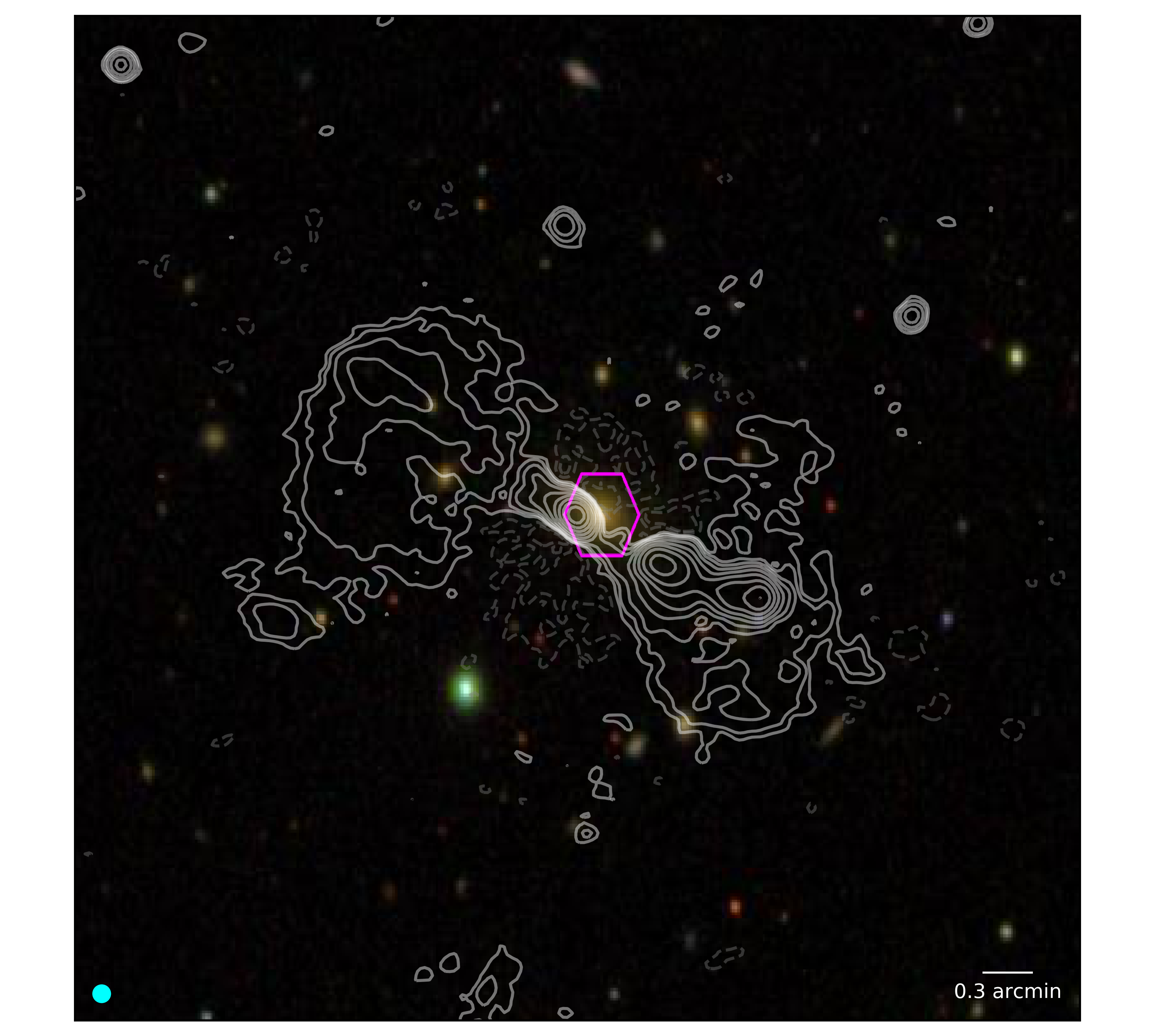}
     \end{subfigure}
        \caption{Overlay of LOFAR 150 MHz radio contours on optical SDSS three color image of late-type RDAGN 8978-9101 and early type RDAGN 8244-6103 (bottom). The magenta hexagon represents the MaNGA IFU footprint. Positive contours are defined by rms noise $\times$ [3, 6, 12, 24, 48, 96, 192, 384, 768, 1536, 3072]. Negative contours are shown by the grey, dashed line and represent the rms noise $\times$ [-3, -6, -12]. The LOFAR beam size is shown in the lower left corner of each image. }
        \label{fig:Galaxies contoured}
\end{figure}

We select our radio-detected AGN (RDAGN) sample by combining the classifications from these four selection techniques to determine an overall classification for each galaxy in the MaNGA-LoTSS catalog. In the diagnostic diagrams presented in Figure \ref{fig:RDAGN Diagnotstic Diagrams}, galaxies can be classified as radio-AGN, SF, intermediate, or unclassified (i.e. low signal-to-noise ratio (S/N) or no measurement), which results in 192 unique combinations of classifications. When choosing the final classification, we weighted each classification from the diagnostic diagrams equally. Galaxies classified as intermediate in the D$_{n}$4000 vs. L$_{1.4\;\mathrm{GHz}}$ / M$_{*}$, [NII] BPT diagrams, or L$_{\mathrm{H\alpha}}$ vs. L$_{150\;\mathrm{MHz}}$ were chosen to ``favor" AGN over SF in order to build the most complete sample of AGN possible. For example, if a galaxy's classification is intermediate in the D$_{n}$4000 vs. L$_{150\;\mathrm{MHz}}$ / M$_{*}$ and [NII] BPT diagrams, SF in the L$_{\mathrm{H\alpha}}$ vs. L$_{150\;\mathrm{MHz}}$, and AGN in the WISE Color-Color diagram, the overall classification of the galaxy is AGN. Any combination that consisted of half SF and half AGN is ``unclassified''. Similarly, a galaxy is unclassified if it has a combination consisting of the following designations: one AGN, one SF, one intermediate, and one unclassified.  We show the overall classifications based on the adopted diagnostic diagrams in Table \ref{tab: Galaxy classification}. Only combinations that classified five or more galaxies are shown to save space. In total, there are 380 AGN galaxies, 783 SF galaxies, and 247 unclassified galaxies. From the 380 AGN galaxies, we removed galaxies that had \verb|MANGA_DRP3QUAL| flags indicating that the final cubes and RSS files did not meet quality standards. Additionally, we visually inspected the radio contours and removed galaxies that had no radio emission greater than 3 $\times$ the rms noise (41 galaxies, see Table \ref{tab:excluded AGN}). Our final RDAGN sample consists of 307 unique RDAGN-host galaxies. In Table \ref{tab:Galaxy classification}, we provide the number of galaxies classified as AGN / SF / intermediate / unclassified in each diagnostic diagram for the entire MaNGA-LoTSS catalog and for the final RDAGN sample. We provide radio-optical overlays of two of the RDAGN host galaxies in Figure \ref{fig:Galaxies contoured}. The example in the top panel of Figure \ref{fig:Galaxies contoured} exhibits radio emission likely powered by both SF and AGN activity. Conversely, the early type galaxy example (bottom panel of Figure \ref{fig:Galaxies contoured}) has two-sided radio jets.

\begin{figure}[!t]
    \centering
    \includegraphics[width=\linewidth, trim=0.75cm 1.1cm 0.75cm 1.1cm, clip]{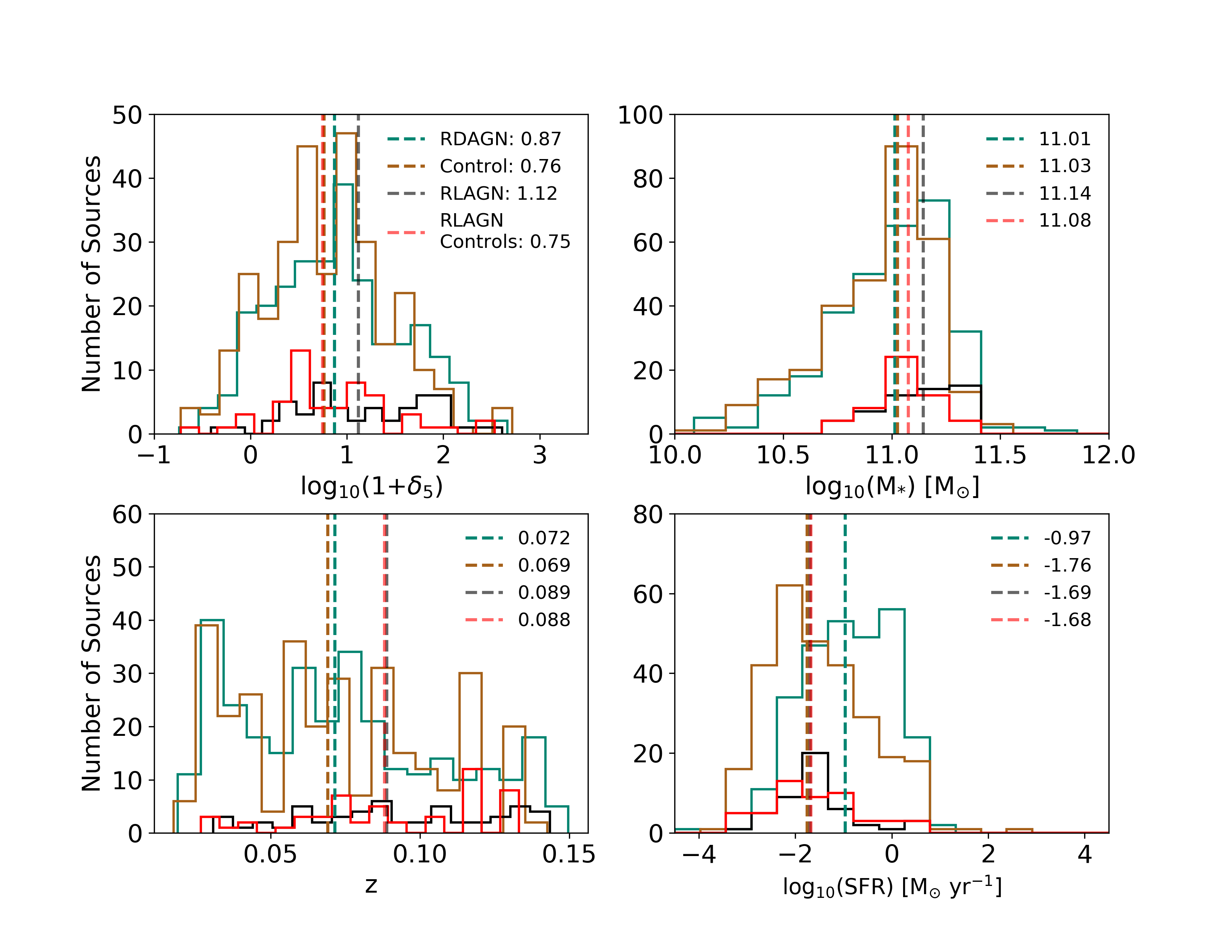}
    \caption[Distribution of measured properties of the RDAGN and control sample.]{Distribution of measured properties of the RDAGN (green), control sample (brown), the classical RLAGN (black), and RLAGN control galaxies (red). The median value for each sample is indicated by the dashed, vertical lines. \textit{Top row from left to right:} local galaxy overdensity evaluated at the fifth nearest neighbour ($\delta_{5}$), M$_{*}$. \textit{Bottom row from left to right:} \textit{z}, SFR as measured by Pipe3D.}
    \label{fig:Histograms}
\end{figure}

\subsection{Control sample criteria}
\label{subsec:Control sample criteria}

From the galaxies in MaNGA DR16 within the LoTSS DR2 footprint, we have selected a control sample of galaxies that closely match the properties of the RDAGN host galaxies except that their nuclei, based on the [NII] BPT and the equivalent width of H$\alpha$ (W(H$\alpha$)) vs. [NII]/H$\alpha$ \citep[WHAN; ][]{CidFernades2010} diagrams, are inactive. The control sample was built as follows:
First, we selected galaxies whose overall classification was not ``AGN” and whose central ionizing source was not AGN. Therefore, we considered a galaxy as a potential control sample candidate if it was in the SF region of the BPT diagram or was classified as a Low-Ionization Emission-line Region (i.e. in the Low-Ionization Nuclear Emission-line Region \citep[LINER; ][]{Heckman1980} region with W(H$\alpha$) $<$ 3 \text{\r{A}}).
From these inactive galaxies, we created a preliminary list of control sample candidates for each RDAGN host, selecting galaxies whose \textit{z} and M$_{*}$ did not vary by more than 30$\%$ from the RDAGN host’s \textit{z} and M$_{*}$. Finally, we selected one control galaxy for each RDAGN host galaxy by visually inspecting the SDSS three color image of each control sample candidate and choosing the galaxy whose morphology and inclination were most similar to the SDSS three color image of the RDAGN host galaxy. Priority was given to morphological features within the MaNGA IFU hexagon footprint. We provide the plateifu identifer for the RDAGN galaxies and their assigned control galaxy in Table \ref{tab:RDAGN and controls}.
\begin{figure}[!t]
    \centering
    \includegraphics[width=0.4\textwidth, trim=0 0.5cm 0 0 clip]{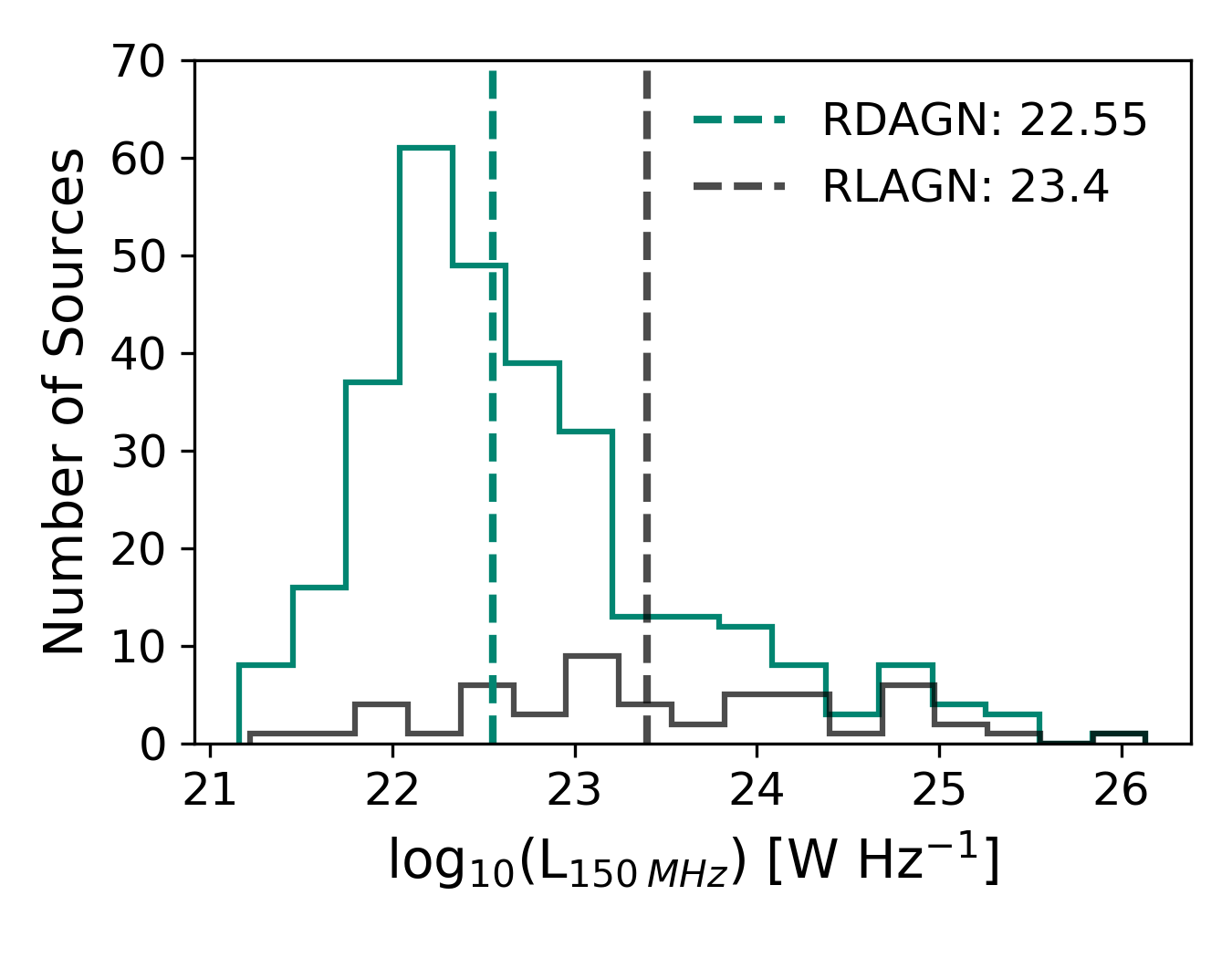}
    \caption[]{Distribution of the L$_{150\;MHz}$ for the RDAGN (green) and RLAGN (grey) samples. The median value is indicated by the dashed, vertical line.}
    \label{fig:L150 Histogram}
\end{figure}

In some cases, a particular MaNGA galaxy was the best control galaxy for multiple RDAGN samples. For example, although we identify 307 RDAGN galaxies, there are only 157 unique controls. Hence, we use the same best control galaxy more than once so that the total number of RDAGN and controls are equal. To ensure that using the same control galaxy multiple times and visually selecting control galaxies did not affect our results, we performed the analyses presented in Sections \ref{section:Relation to the SFMS} and \ref{section:Stellar age gradient} using the entire non-active galaxy sample (3231 galaxies in total), and found that our results did not change. We chose to use our selected control sample to better understand how RDAGN host galaxies compare to non-active galaxies with similar properties and to overcome potential biases in our SFR measurements (see Section \ref{section:Spatially resolved stellar and nebular gas properties}). In Figure \ref{fig:Histograms}, we show the distribution of properties of the RDAGN sample and of the control sample. By selection, the stellar mass and redshift distributions are the same. Moreover, we find no significant difference in the environment in which these RDAGN and control galaxies reside. Our RLAGN span a large range of stellar mass and radio power based on the distribution of L$_{150\;\mathrm{MHz}}$ provided in Figure \ref{fig:L150 Histogram}.

\subsection{Existing MaNGA AGN Catalogs}
\label{subsec:Global properties}

We briefly compare our sample of 307 unique RDAGN to previous studies that have identified AGN in the MaNGA survey, \citep{Rembold2017,Wylezalek2018,Sanchez2018,Comerford2020}. \citet{Rembold2017, Wylezalek2018}, and \citet{Sanchez2018} select AGN using optical emission line ratios and cuts in the EW(H$\alpha$). In our sample of 307 RDAGN, 100 (33\%) have EW(H$\alpha$)$>$1.5 \text{\r{A}}, and 41 (13\%) have EW(H$\alpha$)$>3$ \text{\r{A}}. In the following, we report the percentage of our sample that overlaps with the other MaNGA AGN catalogs and the percentage of our high-EW(H$\alpha$) subsample that overlaps. We note, however, that our sample is distinct from these other MaNGA catalogs with RDAGN \citep[e.g. ][]{Comerford2020} because with LOFAR, we are able to detect fainter radio emission from AGN than previously possible. In Figure \ref{fig:L150 Histogram} we present the distribution of the radio luminosity for our RDAGN sample. The distribution peaks at $\sim$ 22.5 W Hz$^{-1}$, which is lower than the average equivalent 1.4 GHz radio luminosities of radio-AGN in \citet{Best2005b}. These lower luminosities are consistent with the results of \citet{Sabater2019}, who found many RDAGN at the luminosity range 21 $<$ $\mathrm{\log}$(L$_\mathrm{150 MHz}$ [W Hz$^{-1}$]) $<$ 24 that are only detected with the deeper LoTSS data, and not found in NVSS/FIRST.

\begin{figure}[!t]
    \centering
    \includegraphics[width=.47\textwidth, trim=0 0.25cm 0 0, clip]{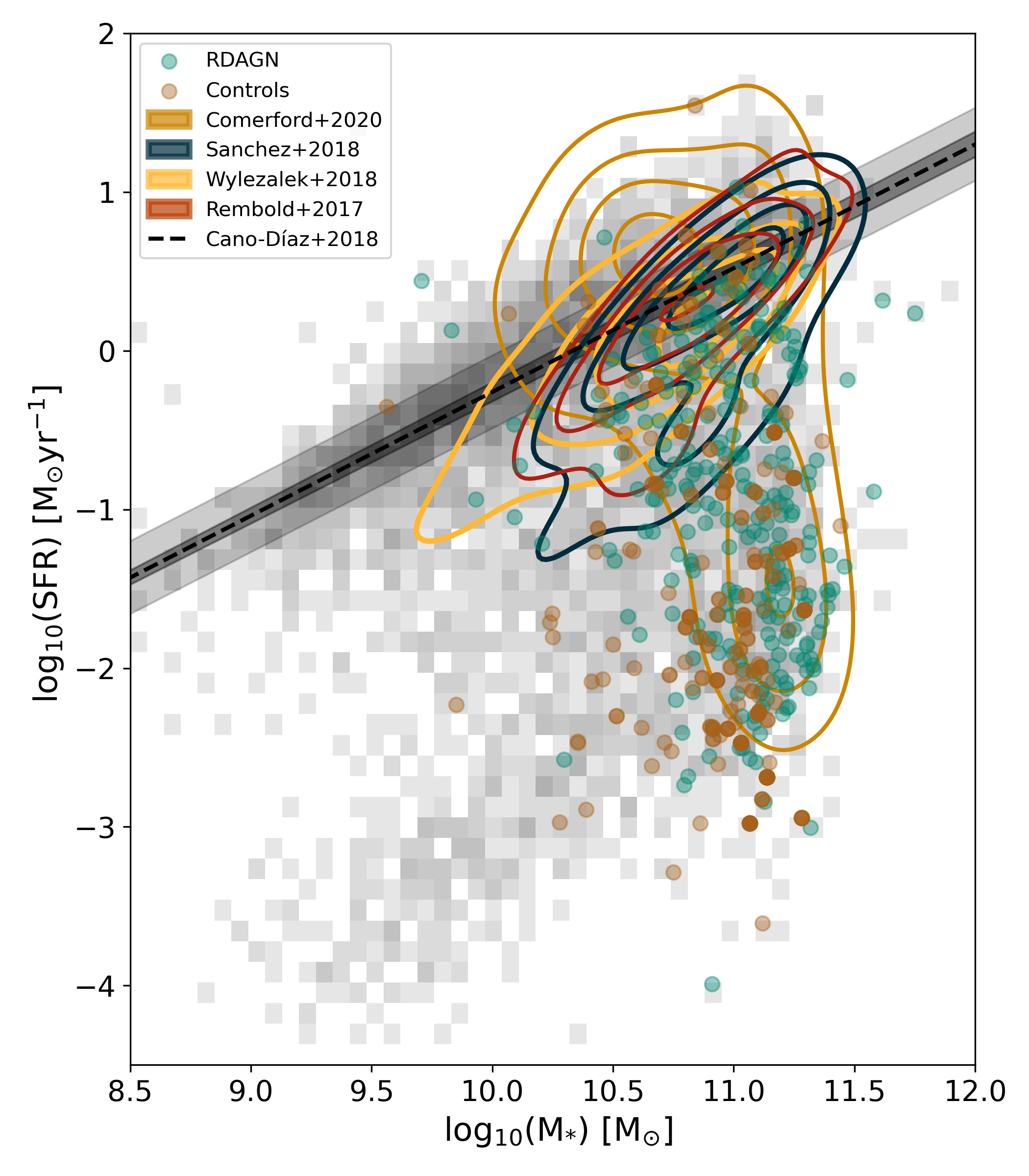}
    \caption[MaNGA AGN Catalogs Relation to SFMS]{Relation between SFR and M$_{*}$ for existing MaNGA AGN catalogs. The grey-colored image represents the density of MaNGA galaxies in the plot. The RDAGN studied in this work are indicated in green. Dark yellow contours represent the density of AGN host galaxies for from the \citet{Comerford2020} AGN catalog, the dark blue contours represent the density of galaxies from the \citet{Sanchez2018} AGN catalog, the \citet{Wylezalek2018} AGN catalog are represented by orange contours, and the red contours exhibit the density of \citet{Rembold2017} AGN catalog. The SFR-M$_{*}$ space is divided into 50x50 bins, and the contours are drawn at 25, 50, 75, and 100$\%$ of the maximum number density. The dotted line represents the SF main-sequence derived for SDSS-IV MaNGA galaxies derived by \citet{Cano-Diaz2019}, the dark grey shading represents the errors on slope, and the light grey shading represents the standard deviation.}
    \label{fig:MaNGA AGN Catalogs Relation to SFMS}
\end{figure}

\subsubsection{\citet{Rembold2017} catalog}

\citet{Rembold2017} used SDSS integrated spectra to construct the [NII] BPT diagram \citep{Baldwin1981} and WHAN diagram to identify ``true" AGN in the galaxies observed in the fifth MaNGA Product Launch (MPL-5). \citet{Rembold2017} identify 62 ``true" AGN out of the 2778 galaxies (2727 unique galaxies) observed in MPL-5. Of the 62 AGN presented in \citet{Rembold2017}, 11 are in our radio-detected AGN catalog ($\sim$ 4\%; 27\% with EW(H$\alpha$)$>$3\,\text{\r{A}}).

\subsubsection{\citet{Wylezalek2018} catalog}

\citet{Wylezalek2018} used spatially resolved methods to identify AGN candidates in MPL-5 and determined the classification of each spaxel based on its location on the [NII] and [SII] BPT diagrams. Their sample consists of 308 ``AGN candidates" that have a high spaxel fraction of AGN in both the [NII] and [SII] BPT diagrams and have cuts on the equivalent width and surface brightness of H$\alpha$. 28 AGN candidates from \citet{Wylezalek2018} are in our AGN catalog ($\sim$ 9\%; 32\% with EW(H$\alpha$)$>$3\,\text{\r{A}}).

\subsubsection{\citet{Sanchez2018} catalog}

\citet{Sanchez2018} chose AGN using Pipe3D’s integrated emission-line ratios within the central 3$^{\prime\prime}\times$3$^{\prime\prime}$ of MPL-5 galaxies. They classified galaxies as AGN if their integrated emission-line ratios were above the [NII] BPT maximum starburst line from \citet{Kewley2001} and whose W(H$\alpha$) was greater than 1.5\text{\r{A}}. \citet{Sanchez2018} identified 98 AGN from the 2700 galaxies in MPL-5, 22 of which overlap with our RDAGN sample ($\sim$ 7$\%$; 19\% with EW(H$\alpha$)$>$1.5\,\text{\r{A}}).

\subsubsection{\citet{Comerford2020} catalog}

\citet{Comerford2020} selected AGN in galaxies observed in MPL-8 using broad Balmer emission lines from SDSS DR7 spectra, radio observations from NVSS \citep{Condon1998} and FIRST \citep{Becker1995}, WISE mid-infrared colors, and ultra-hard X-ray observations from the Swift observatory’s Burst Alert Telescope (BAT). \citet{Comerford2020} used the SDSS DR7 AGN catalog from \citet{Best2012} to identify radio AGN in MaNGA MPL-8. \citet{Best2012} selected radio AGN using the D$_{n}$4000 vs. L$_{1.4\;\mathrm{GHz}}$ / M$_{*}$, [NII] BPT diagram, and L$_{\mathrm{H\alpha}}$ vs. L$_{150\;\mathrm{MHz}}$. Unlike the WISE diagnostic methods presented in this study, \citet{Comerford2020} follow \citet{Assef2018}, adopting the 75$\%$ reliability criteria of \(W1-W2 > 0.486e^{0.092(W2-13.07)^2}\) and W2 $>$ 13.07, or W1-W2 $>$ 0.486 and W2 $\leq$ 13.07. Of the 6261 galaxies observed in MPL-8, \citet{Comerford2020} identify 406 unique AGN. \citet{Comerford2020} focused their analyses on comparing 81 radio-quiet galaxies undetected in the radio with 143 radio-mode AGN. 52 AGN from \citet{Comerford2020} are in our AGN catalog ($\sim$17\% of our sample, or $\sim$ 13$\%$ of their total AGN and 38$\%$ of their RDAGN sample).

\subsection{Comparison of global properties}

We show SFR as a function of M$_{*}$ for our sample of RDAGN, as well as for those AGN in other MaNGA AGN catalogs outlined above in Figure \ref{fig:MaNGA AGN Catalogs Relation to SFMS}. Compared to optically selected AGN catalogs \citep[][]{Rembold2017,Sanchez2018,Wylezalek2018}, radio-selected AGN \citep[][and this study]{Comerford2020} have a strong tendency to occupy massive host galaxies. This is already a well observed trend \citep[e.g. ][]{Gurkan2018, Sabater2019} and indicates that radio-AGN selection intrinsically selects for a different population of host galaxies (i.e. massive early-type galaxies -- hereafter ETGs -- for radio-AGN, and less massive late-type galaxies -- hereafter LTGs -- for optically selected AGN). Figure \ref{fig:MaNGA AGN Catalogs Relation to SFMS} illustrates that galaxies experiencing quenching (Green Valley galaxies) and AGN host galaxies share a similar location on the SFR$_{*}$ plane, which is also well observed \citep[e.g. ][]{Sanchez2018, Lacerda2020}.

\section{Relation to the star-forming main-sequence (SFMS)}
\label{section:Relation to the SFMS}
\begin{figure*}[!th]
    \centering
    \includegraphics[width = \textwidth, clip]{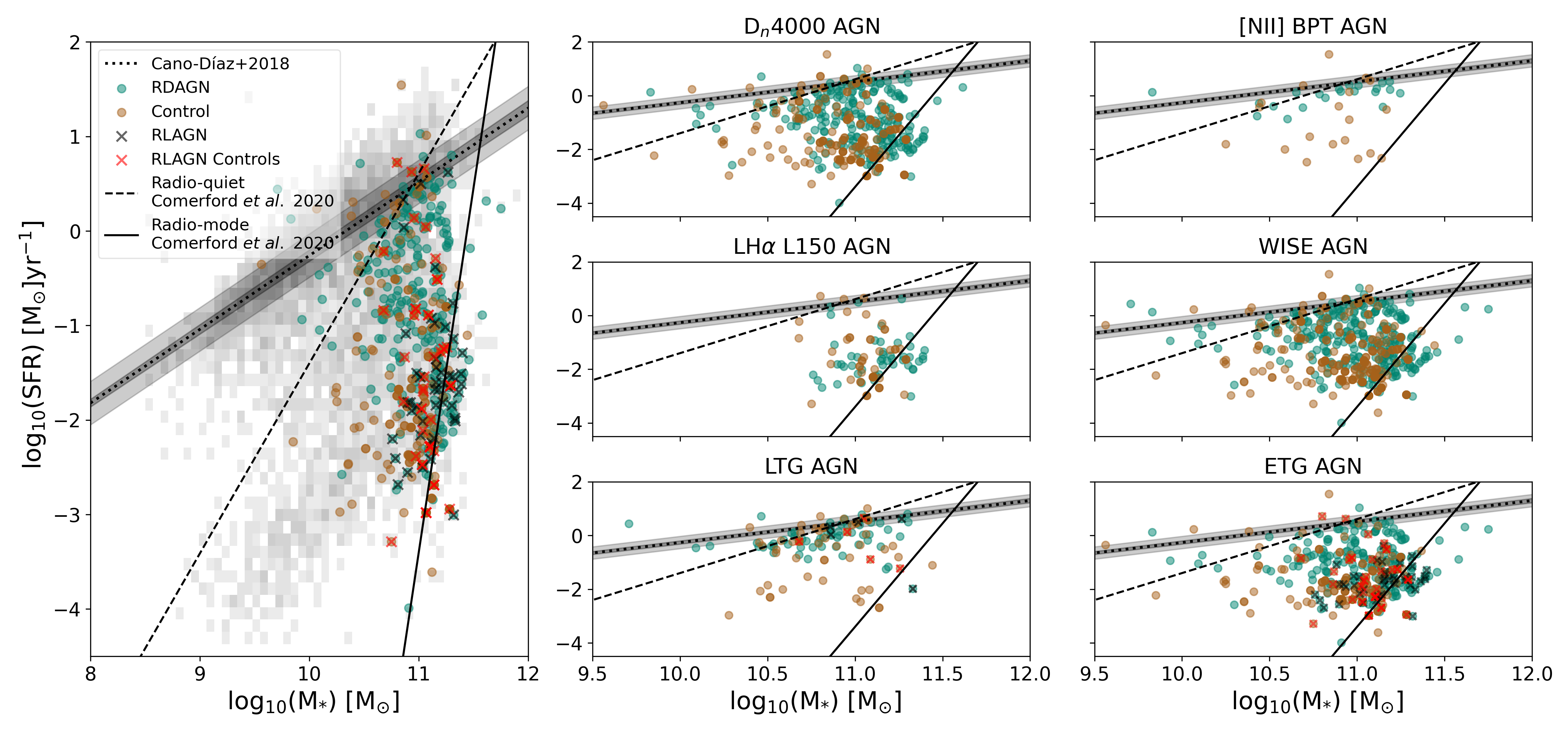}
    \caption[SFR vs. M$_{*}$ for the full RDAGN sample and control sample.]{Relation between SFR and M$_{*}$ for RDAGN host galaxies (green), the full control sample (brown), RLAGN sub-sample (black x's), RLAGN control galaxies (red x's), and entire Pipe3D catalog (grey). For reference, the SF main-sequence derived for SDSS-IV MaNGA galaxies is indicated by the dotted line \citep{Cano-Diaz2019}, the dark grey shading represents the errors on slope, and the light grey shading represents the standard deviation. The best-fit relations for radio-quiet AGN (dashed line) and radio-mode AGN (solid line) are from \citet{Comerford2020}.}
    \label{fig:SFR vs. SM Full}
\end{figure*}

\begin{figure*}[!th]
    \centering
    \includegraphics[width = \textwidth, clip]{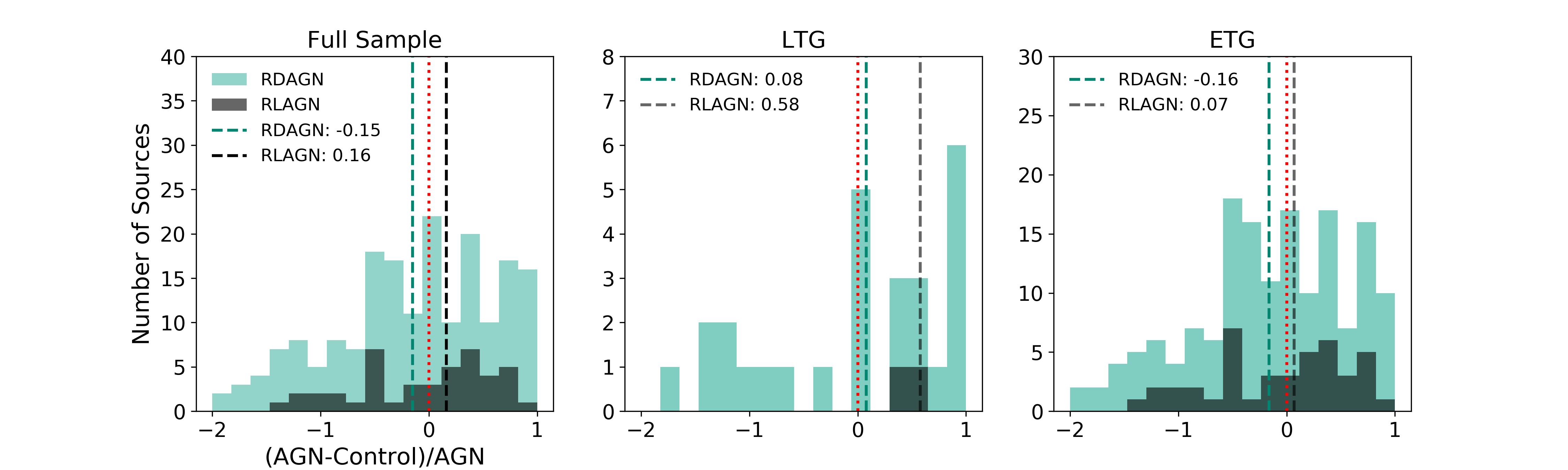}
    \caption[]{Distribution for the fractional difference in the SFR as measured by Pipe3D of the RDAGN and its control galaxy for the entire sample, LTGs, and ETGs (green shading). The same values are also shown for the classical RLAGN sub-sample and its controls (black shading). The vertical, dashed lines represent the median of the distribution. A one-to-one line at zero is represented by the red, dotted line.}
    \label{fig:FracDiff}
\end{figure*}

Star-forming galaxies fall on a tight correlation \citep[$\sim$ 0.2 intrinsic scatter; ][]{Speagle2014} between SFR and M$_{*}$, known as the main-sequence \citep{Brinchmann2004, Daddi2007, Elbaz2007, Noeske2007}. Previous studies have found that AGN add additional complexity to the regulatory processes of SF when compared to non-active galaxies of comparable mass. For example, \citet{Mullaney2015} and \citet{Shimizu2015} have found that the SFRs in X-ray-selected AGN host galaxies have more suppressed SFRs than non-active galaxies of similar mass, whereas \citet{Young2014} and \citet{Pitchford2016} have found that quasar host galaxies have higher SFRs than comparably massive, non-active galaxies. Moreover, the type of AGN -- such a radio-quiet/radio-loud, low excitation radio galaxies (LERGs) / high excitation radio galaxies (HERGs) -- that a galaxy hosts appears to influence the SFRs \citep[e.g. ][]{Hardcastle2013, Heckman2014, Ellison2016, Magliocchetti2016, Magliocchetti2018, Roy2018, Comerford2020}.

In this section, we show the relation of RDAGN host galaxies and control galaxies to the SFMS using the integrated stellar mass and SFR from the Pipe3D VAC. In Figure \ref{fig:SFR vs. SM Full}, we show the correlation between SFR and M$_{*}$ for the full sample and for AGN sub-samples based on the diagnostic diagrams used in Section \ref{subsec:Selecting radio-detected AGN} and based on morphology. We include the star-forming main-sequence (SFMS) relation derived for SDSS-IV MaNGA galaxies from \citet{Cano-Diaz2019}, which is defined by \(\log(\mathrm{SFR}/M_{\odot}yr^{-1}) = -8.06 \pm 0.04 + (0.78 \pm 0.01) \times \log \mathrm{M_{\*}}/M_{\odot})\), and has a standard deviation of 0.23. Additionally, we show the best-fit relation for radio-quiet and radio-mode MaNGA AGN derived by \citet{Comerford2020}. These lines are defined as \(\log(\mathrm{SFR}/M_{\odot}yr^{-1}) = \alpha + \beta \log(M_{*}/M_{\odot})\), where $\alpha = -$ 88.1 $\pm$ 8.1 and $\beta = $7.7 $\pm$ 0.7 for radio-mode AGN and $\alpha = -$ 21.5 $\pm$ 0.7 and $\beta = $2.01 $\pm$ 0.06 for radio-quiet AGN.

We present the M$_{*}$-SFR function for our RDAGN and control samples in Figure \ref{fig:SFR vs. SM Full} and find that both the RDAGN sample and the control sample typically lie below the main-sequence. To confirm the observed similarity between the two samples, we calculated the distance from the SFMS ($\Delta$ log$_{10}$(SFR)) by subtracting the (logarithmic) SFR of the SFMS from the SFR of the sample (values presented in Table \ref{tab:Pipe3D Results}). Although the median $\Delta$ log$_{10}$(SFR) of the RDAGN sample (-1.51 dex $\pm$ 3.20) lies closer to the SFMS than the median of the control sample ($\Delta$ log$_{10}$(SFR)= -2.29 dex $\pm$ 3.03), the standard deviation errors on the median overlap. Therefore, the difference between the median $\Delta$ log$_{10}$(SFR) at a fixed stellar mass for the RDAGN sample and the control sample is not statistically significant (see Table \ref{tab:Pipe3D Results}). 

RDAGN classified as ``AGN" in the [NII] BPT diagram and those residing in LTGs tend to agree with the best-fit relation for radio-quiet AGN of \citet{Comerford2020} (median $\Delta$ log$_{10}$(SFR) $\sim$ -0.359 and -0.465, respectively). This is expected as the BPT diagram tends to select radiatively efficient AGN, which are typically radio quiet. Furthermore, radio quiet AGN are often hosted by LTGs. 

Conversely, we find that early-type RDAGN and RDAGN classified as ``AGN" on the L$_{\mathrm{H\alpha}}$ vs. L$_{150\;\mathrm{MHz}}$ typically agree with the best fit relation for radio-mode AGN \citep[][median $\Delta$ log$_{10}$(SFR) $\sim$ -1.74 and -2.35, respectively]{Comerford2020}. This is again expected as the selection criterion L$_{\mathrm{H\alpha}}$ vs. L$_{150\;\mathrm{MHz}}$ selects radio loud objects, and radio-loud AGN  typically reside in ETGs. 

\begin{figure*}[!ht]
    \centering
    \includegraphics[width = 0.78\textwidth, trim = 0 1.cm 0 1.15cm, clip]{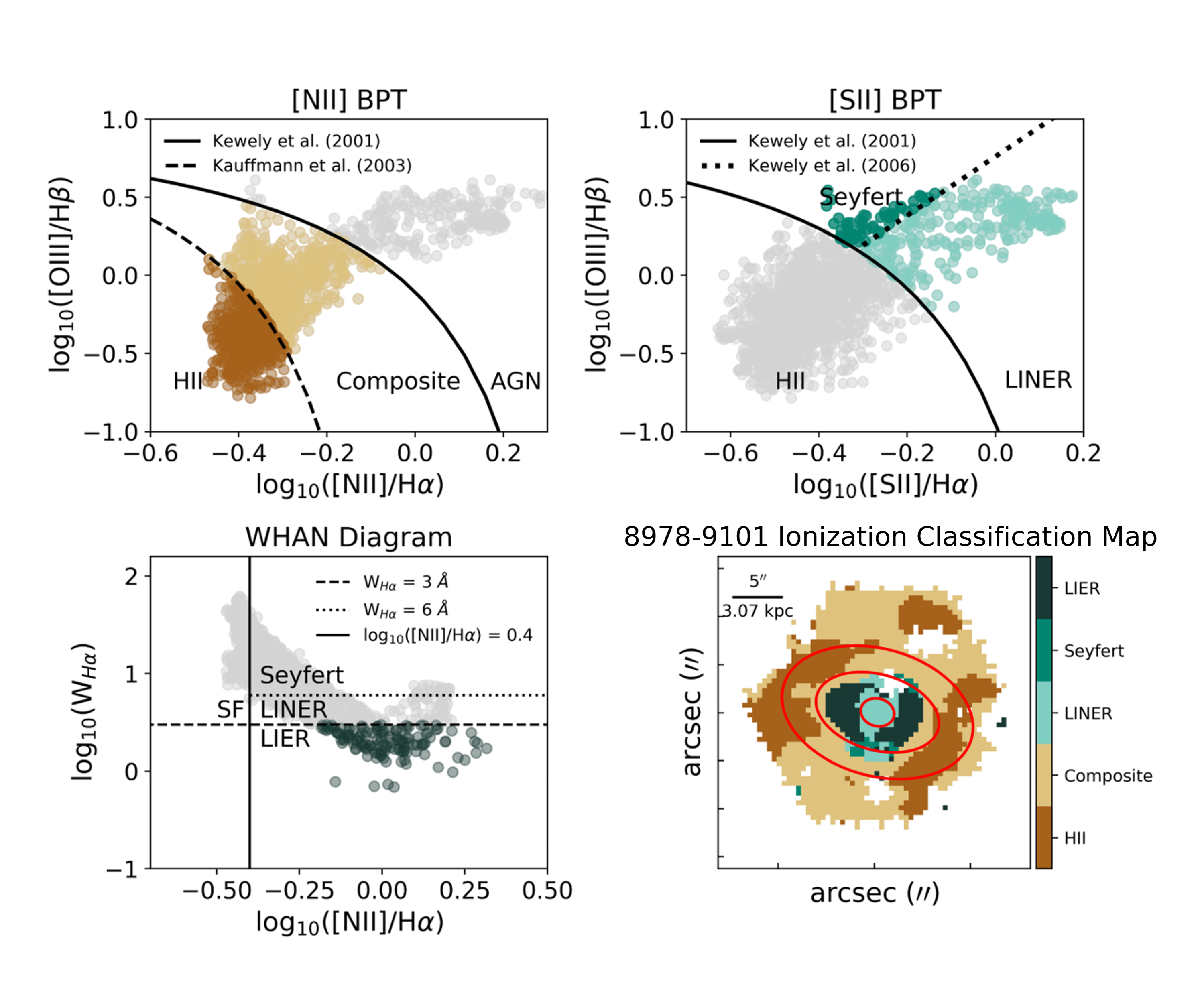}
    \caption[Excitation map construction.]{Example of the diagrams used to determine the gas excitation mechanisms across the surface of each galaxy. Each point on the diagrams represents a spaxel. \textit{Top row from left to right: }[NII] BPT diagram and [SII] BPT diagram of a late-type RDAGN galaxy example. The solid line represents the maximum starburst line from \citet{Kewley2001}. The dashed line on the [NII] BPT diagram represents the \citet{Kauffmann2003} line, which separates pure SF galaxies from composite galaxies. The dotted line on the [SII] BPT diagram separates Seyfert-like excitation from LINER-like excitation \citep{Kewley2006}. \textit{Bottom row from left to right: }WHAN diagram and final ionization classification map of RDAGN galaxy 8978-9101. The [NII] BPT was used to separate HII (brown) and composite (beige) excitation, the [SII] diagram was used to distinguish Seyfert-like excitation (green) from LINER-like excitation (light green), and the WHAN diagram was used to differentiate LIER-like excitation (dark green) from LINER-like excitation. The red ellipses represent, from the inside outwards, 0.2 R/R$_{e}$, 0.6 R/R$_{e}$, 1 R/R$_{e}$.}
    \label{fig:Constructing excitation map}
\end{figure*}

We have found that the majority of the RDAGN lies below the SFMS, which is consistent with what is expected for the position of radio AGN relative to the main sequence \cite[e.g. ][]{Gurkan2018}. Unlike previous studies \citep[e.g. ][]{Young2014, Mullaney2015, Shimizu2015, Leslie2016, Pitchford2016}, which found that AGN host galaxies have different SFRs than non-active galaxies of similar mass, we find no statistically significant difference between the SFR of the RDAGN sample and the control sample selected by mass and morphology. This result compliments the findings presented in previous explorations of AGN feedback with MaNGA \cite[e.g. ][]{Sanchez2018} and with the CALIFA survey \cite[CALIFA survey (Calar Alto Legacy Integral Field Area; e.g. ][]{Lacerda2020}, specifically that there is no significant difference between the properties of galaxies in the Green Valley hosting an AGN and those without an AGN. Our results indicate that the RDAGN, selected based on their current activity, are not responsible for any quenching that has taken place in their host galaxies. The mechanism or mechanisms responsible for suppressing SF must be related to the host galaxy's properties (i.e. the fact that these are preferentially ETGs, with lower SF than star forming galaxies), which is in agreement with the burgeoning literature that the growth of galactic bulges, AGN activity, and the halting of SF appear to occur concomitantly \cite[e.g. ][and references therein]{Lacerda2020}.

\subsection{Fractional difference of SFRs}

Towards understanding how the SFRs between each RDAGN and its assigned control galaxy directly compare, we look at the fractional difference of Pipe3D's SFR measurement, which is the difference between the SFR of the RDAGN and its control divided by the SFR of the RDAGN ((SFR$_{\mathrm{AGN}} -$ SFR$_{\mathrm{Control}}$)/SFR$_{\mathrm{AGN}}$). Dividing the difference by the SFR of the RDAGN helps scale the range of measured SFRs. When the fractional difference is positive, it means that the RDAGN host galaxy has a higher SFR than its assigned control galaxy. Conversely, when the difference is negative, the control galaxy has a higher SFR. We present the distribution of the fractional difference in Figure \ref{fig:FracDiff}.

In Figure \ref{fig:FracDiff}, the fractional difference of the SFR between the RDAGN sample and control sample is represented by the distribution shaded in green. We find that $\sim$ 44$\%$ of the RDAGN-control pairs exhibit a positive fractional difference. The percentage increases when late-type AGN host galaxies are considered; $\sim$ 51$\%$ of the RDAGN LTGs have higher SFRs than the corresponding controls. Finally, for the ETGs, only $\sim$ 43$\%$ of the RDAGNs have higher SFRs. For classical RLAGN and their corresponding control galaxies, we discover higher percentage of positive fractional differences. The full RLAGN sample and the early-type RLAGN sub-sample express a similar percentage ($\sim$ 54$\%$) of positive fractional differences. We find that $\sim$ 80$\%$ of late-type RLAGN express a positive fractional difference.

Our fractional difference of SFR results are both agree and disagree with those of \citet{doNascimento2019} (a MaNGA AGN study that uses the \citet{Rembold2017} catalog). We note that the SFR measurements that \citet{doNascimento2019} use in their fractional difference analysis were taken using similar methods outlined in Section \ref{subsec:SFR surface denisty}. We chose to use Pipe3D's values instead of the ones we calculate in Section \ref{subsec:SFR surface denisty} in order to have a SFR measurement for each RDAGN and control galaxy (discussed further at the beginning of Section \ref{section:Spatially resolved stellar and nebular gas properties}). Nevertheless, both \citet{doNascimento2019} and Pipe3D measure SFR using the extinction-corrected L$_{\mathrm{H\alpha}}$ equation from \citet{Kennicutt1998} (see Equation \ref{eq: Kennitcutt SFR}) facilitating comparison.

Whereas \citet{doNascimento2019} find that 76$\%$ of ETG AGN have higher SFRs than their assigned control galaxies, only $\sim$ 43$\%$ of our RDAGN ETG host galaxies have higher total SFRs than their controls. Our values agree more when comparing the percentage of positive fractional differences in the early-type RLAGN sample ($\sim$ 54$\%$). We believe that the difference in our percentages and those reported by \citet{doNascimento2019} is due the differences in our AGN samples.

Interestingly, we discover that $\sim$ 51$\%$ of our late-type RDAGN host galaxies have higher total SFRs than their controls, which is the same percentage reported by \citet{doNascimento2019}. This might be a sign of either positive feedback playing a role at earlier stages of a galaxy's evolution or that LTGs simply have more availability of fuel. To distinguish between these two scenarios, we would need to prove that radio jet activity is physically reaching regions where SF is occurring.

\section{Spatially resolved stellar and nebular gas properties}
\label{section:Spatially resolved stellar and nebular gas properties}

\subsection{Ionization classification maps}
\label{subsec:Excitation maps}

A galaxy's spectrum contains a wealth of information that is used to infer the physical processes taking place within the galaxy. Historically, the dominant excitation mechanism of a galaxy was inferred using single-aperture spectroscopy \citep[e.g.][and references therein]{Kauffmann2003, Kewley2006}. However, with IFS data, multiple ionizing sources can be determined and spatially mapped because a spectrum of light is measured at every spatial pixel observed with the IFU. Here, we optically classify the spaxels of the RDAGN and control galaxies to separate multiple ionizing sources and to gauge the frequency of these mechanisms at three different galactocentric radii. Knowing where the gas is being excited by these mechanisms is important for obtaining accurate SFR from the luminosity of H$\alpha$, which is the approach used in Section \ref{subsec:SFR surface denisty}.   

\begin{figure}[!ht]
     \centering
     \begin{subfigure}[b]{0.43\textwidth}
         \centering
         \includegraphics[width=\textwidth, trim=0 0.75cm 0 0.75cm]{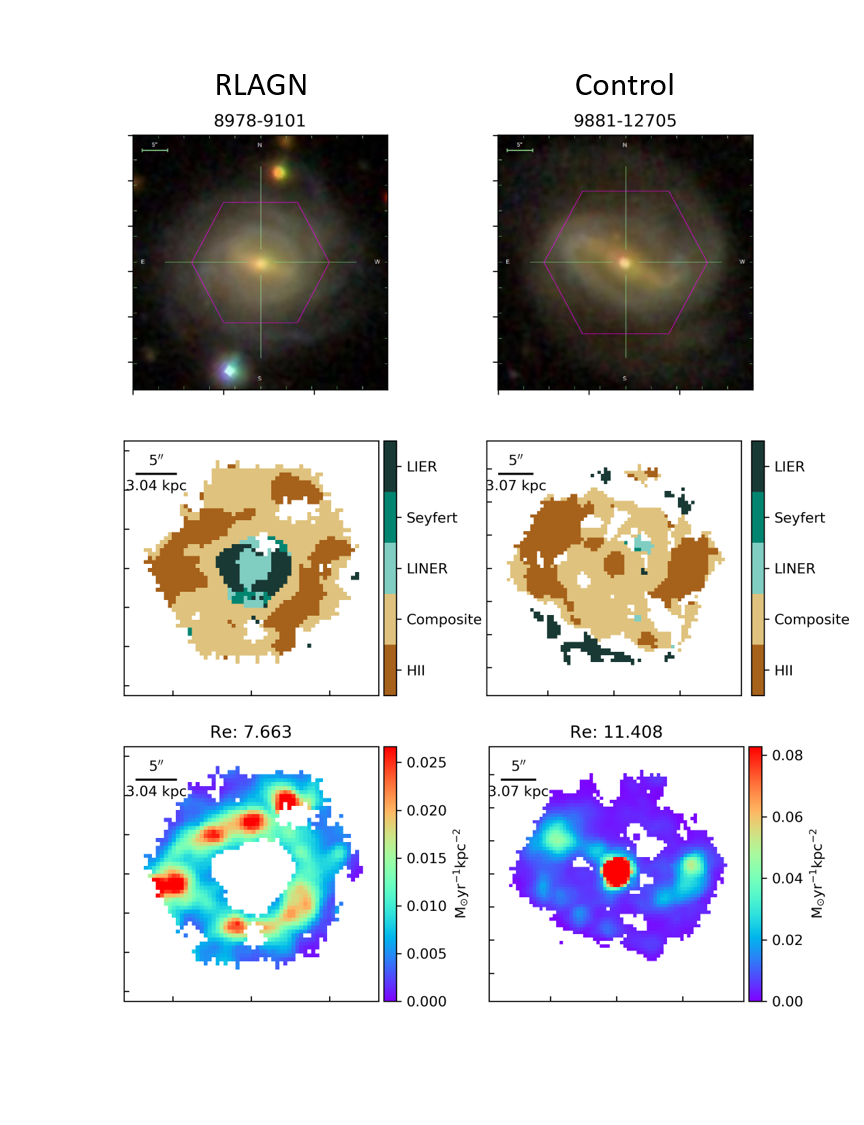}
     \end{subfigure}
     \vfill
     \begin{subfigure}[b]{0.43\textwidth}
         \centering
         \includegraphics[width=\textwidth, trim=0 0.75cm 0 0.75cm]{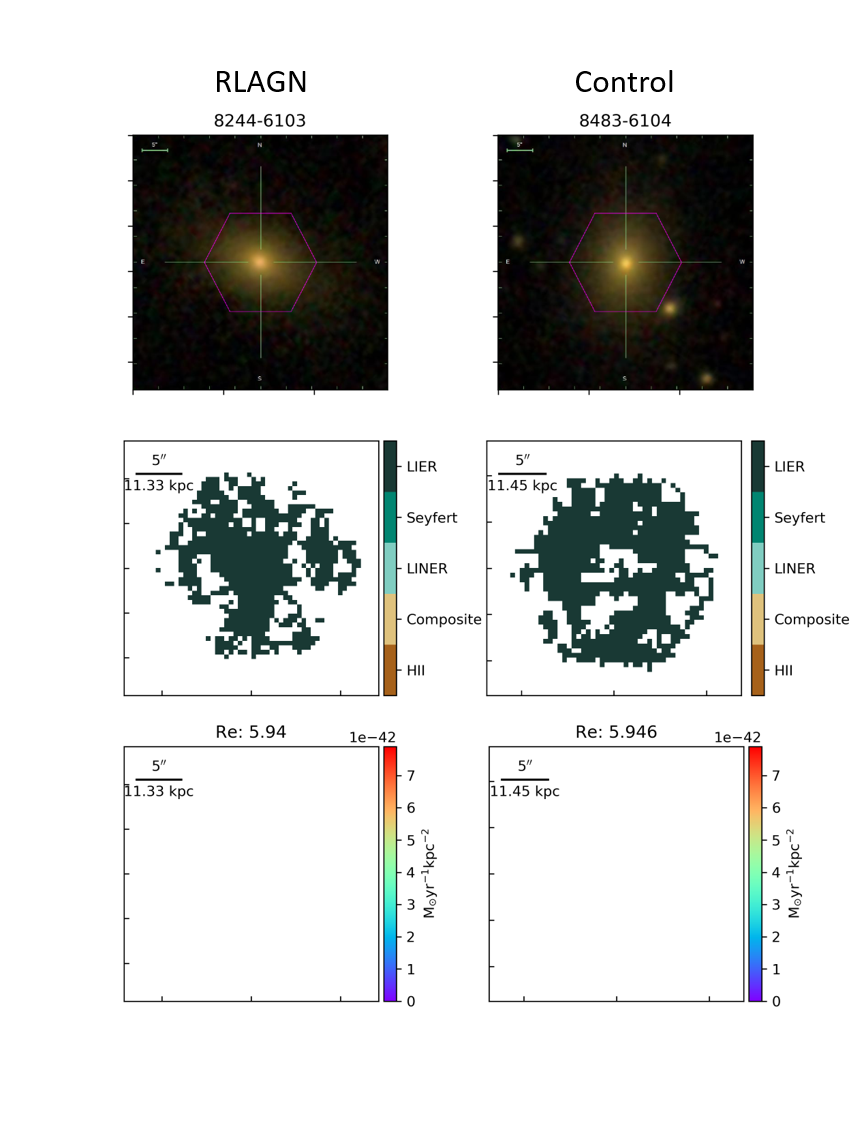}
     \end{subfigure}
        \caption{Surface distribution of gas excitation mechanisms (middle panel) and $\sum$SFR (bottom panel) for late-type RDAGN 8978-9101 and its control galaxy 9881-12705 (top) and early-type RDAGN 8244-6103 and its control galaxy 8483-6104 (bottom). For the early-type RDAGN and its control, the $\sum$SFR maps are blank because those galaxies do not contain SF or composite spaxels. The optical SDSS image overlaid with the MaNGA IFU footprint (magenta hexagon) is shown on the top panel of the figure.}
        \label{fig:spatial maps}
\end{figure}

Emission-line fluxes across the surface for each galaxy were obtained using the \verb|Maps| galaxy tool from SDSS Marvin \citep{Cherinka2019}. We determined the S/N of each 2D map using the \verb|get_snr()| function. We also masked spaxels at six wavelengths (H$\beta $ $\lambda$4862, [OIII] $\lambda$5008, [NII] $\lambda$6585, H$\alpha$ $\lambda$6564, [SII] $\lambda$6718, 6732) that contained negative flux values as well as those that had a S/N less than 3 using the \verb|get_masked()| function. In addition to the emission-line fluxes, we obtained the equivalent width of H$\alpha$ line (W(H$\alpha$)) to construct the W(H$\alpha$) vs. [NII]/H$\alpha$ (WHAN) diagram and measurements of the elliptical radius in order to determine the excitation mechanisms within the nuclear region of each galaxy. 

    To determine the excitation mechanism of each spaxel we combined information obtained from three diagrams: the [NII] BPT diagram, [SII] BPT diagram, and the WHAN diagram. Figure \ref{fig:Constructing excitation map} shows an example of the classification methods combined to create the ionization classification map (bottom right) for RDAGN galaxy, 8978-9101. Using the emission-line fluxes at each spaxel, we measured the ratio of [NII] to H$\alpha$, [SII] to H$\alpha$, and [OIII] to H$\beta$. To determine whether the excitation mechanism was from starburst activity / young hot stars (HII) or ``composite", meaning that the gas is likely being excited by a blend of AGN activity and SF,  we used the [NII] BPT diagram and its diagnostic lines. In Figure \ref{fig:Constructing excitation map}, the [NII] BPT diagram is shown in the upper left panel. The points colored brown and beige represent the spaxels of RDAGN 8978-9101 whose excitation mechanism is SF and composite, respectively. The points colored grey, represent spaxels whose emission is likely powered by AGN activity. 

The [SII] BPT diagram was used to distinguish emission line regions dominated by Seyfert-like and LINER-like excitation. We chose the [SII] BPT (upper right corner of Figure \ref{fig:Constructing excitation map}) to separate these ionizing mechanisms because the low ionization potential of the [SII]/H$\alpha$ reveals the low ionization emission lines of LINER spectra better than [NII]/H$\alpha$. Consequently, the Seyfert-LINER demarcation is more robust on the [SII] BPT than on the [NII] BPT \citep{Kewley2006}. The solid line on the [SII] BPT represents the demarcation between HII excitation from AGN excitation and it is defined by $\log$([OIII]/H$\beta$) $=$ 0.72 / ($\log$([SII]/H$\alpha$) $-$ 0.32) $+$ 1.30 \citep{Kewley2001}, where every spaxel above the line is dominated by AGN activity and SF below the line. Seyfert-like excitation is separated from LINER-like excitation by the line $\log$([OIII]/H$\beta$) $=$ 1.89$\times$ $\log$([SII]/H$\alpha$) $+$ 0.76 \citep{Kewley2006}, which is shown by the dotted line on the [SII] BPT diagram in Figure \ref{fig:Constructing excitation map}. All spaxels that fall above this line are classified as Seyfert and spaxels are classified as LINER if they are below the line. In the [SII] BPT, spaxels whose excitation mechanism is Seyfert-like are colored green and those spaxels with LINER-like excitation are colored light green. 

There are multiple ionizing mechanisms that are connected to LINER-like emission in galaxies. Those mechanisms include shock ionization, a weak AGN, or photo-ionization from hot, evolved stars \citep[e.g. post-asymptotic giant branch stars (pAGB); ][and references therein]{Binette1994,Stasinska2006,Sarzi2010,CidFernandes2011,Yan2012,Belfiore2016}. IFU surveys such as CALIFA and MaNGA have revealed that LINER-like emission-line ratios can be seen throughout galaxies  \citep[e.g. ][]{Singh2013, Belfiore2016}, which is attributed the extended LINER-like emission from pAGB stars, which is known as Low-Ionization Emission line Region-like (LIER) excitation \citep[see ][for a more detailed exploration of the pAGB origin of diffuse ionization in galaxies]{Gomes2016, Lacerda2018, Espinosa-Ponce2020}. To separate LIER-like excitation from LINER-like excitation, we constructed the WHAN diagram (lower left panel of Figure \ref{fig:Constructing excitation map}). Several lines of demarcation appear on the WHAN diagram: the solid, vertical line at log$_{10}$([NII]/H$\alpha$) $=$ -0.4 separates SF (left) from AGN / non-SF activity (right), the dotted, horizontal line separates Seyfert-like excitation from LINER-like excitation, and the dashed, horizontal line separates LIER-like excitation from LINER-like excitation. Points colored dark green on the WHAN diagram in Figure \ref{fig:Constructing excitation map} represent the spaxels in RDAGN 8978-9101 with LIER-like excitation.

After the dominant ionizing mechanism was determined for each spaxel, we spatially mapped (see lower right panel in Figure \ref{fig:Constructing excitation map}) the excitation mechanisms. The ionization classification maps for RDAGN 8978-9101 and RDAGN 8244-6103, are compared to those of their controls in Figure \ref{fig:spatial maps}. In the LTG example (RDAGN 8978-9101, top panel of Figure \ref{fig:spatial maps}), both the RDAGN host galaxy and the control are dominated by spaxels consistent with HII excitation (brown) and by composite emission (beige). In the central 5$^{\prime\prime}$, there is LIER (dark green) and LINER-like (light green) excitation, likely from pAGB stars and from a weak AGN, respectively. Conversely, the spaxels in the ETG example (RDAGN 8244-6103, see Figure \ref{fig:spatial maps}) are mostly classified as LIER (dark green), which likely correspond to their old stellar populations. 

It is important to emphasise that although we have separated ``HII'' and ``Composite'' spaxels, in IFS data, gas with both HII and composite emission line ratios is most likely excited by star-formation. This is why, in Section \ref{subsec:SFR surface denisty}, we calculate SFRs from the Balmer lines in both HII and composite spaxels. We should also keep in mind that shocks can reproduce line ratios that are typical for the HII, Composite, to the Seyfert and LINER regions of the diagnostic diagrams (e.g. \citealt{Allen2008}). Future work to identify shocks from mergers or outflows driven by star-formation or AGN activity in our sample will require a combination of emission line analysis with spatial and velocity information \citep[e.g. ][]{LopezCoba2019, LopezCoba2020}.

\begin{figure}[!t]
    \centering
    \includegraphics[width = 0.49\textwidth, trim=0 0cm 0 0cm]{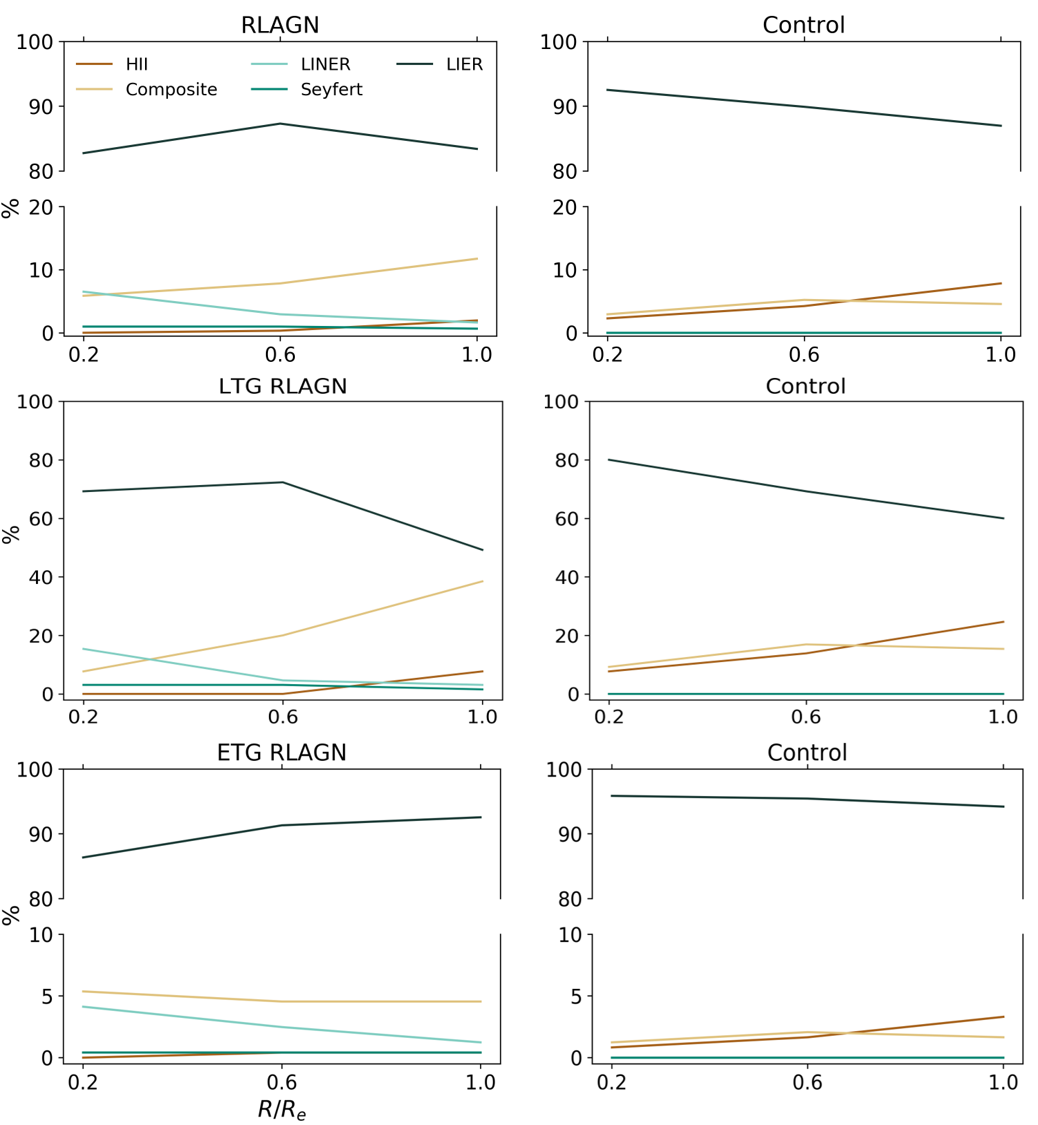}
    \caption[Percentage of galaxies with SF, Composite, LINER, Seyfert, and LIER-like excitation within  0.2, 0.6, 1.0 effective radius (R$_{e}$) from the nucleus of the galaxy.]{Percentage of galaxies that have emission typical of SF, Composite, LINER, Seyfert, and LIER activity within  0.2, 0.6, and 1.0 effective radius (R$_{e}$) from the nucleus of the galaxy. We use a broken y-axis for the top and bottom rows.}
    \label{fig:Excitation bar chart}
\end{figure}

In Figure \ref{fig:Excitation bar chart} we provide line graphs, which display the percentage of galaxies that have HII, Composite, LINER, Seyfert, and LIER at 0.2, 0.6, and 1.0 effective radius (R$_{e}$) as the dominant excitation mechanism and provide the numerical values in Table \ref{tab: spaxel percentages}. Before elaborating further on these results, some samples appear to not have certain spaxel-types (i.e. 0$\%$). To be clear, that does not mean that the specific excitation mechanism does not occur in that given galaxy. Instead, it means that the excitation type was not the dominant ionizing mechanism (i.e. by number of spaxels) within the radial bin of 0.2, 0.6, or 1.0 R$_{e}$.

We find that within 0.2 R$_{e}$ of each galaxy (i.e. the nuclear region), LIER-like excitation (represented by the dark green line in Figure \ref{fig:Excitation bar chart}) is the most common ionizing mechanism in all samples. $\sim$85$\%$ of RDAGN galaxies and $\sim$93$\%$ of control galaxies exhibit LIER spaxels near the nuclear region. Approximately 69$\%$ and 86$\%$ of LTGs and ETGs galaxies are dominated by as LIER spaxels in the nuclear region, respectively. At larger effective radii, LIER spaxels become less common (varies between $\sim$83-85$\%$ for the entire RDAGN sample and $\sim$87-92$\%$ for the control sample), but still remain the dominant excitation mechanism. In the RDAGN sample and the sub-sample of early type AGN host galaxies, the percentage of galaxies dominated by LIER spaxels peaks at 0.6 R/R$_{e}$ ($\sim$87$\%$). The presence of LIER-like emission throughout the entire galaxy, regardless of activity or morphology, is consistent with previous studies\citep[e.g. ][]{Singh2013, Gomes2016, Belfiore2016, Wylezalek2018}. Although pAGB stars are likely responsible for the photoionziation of gas in these spaxels, another possible interpretation of the LIER emission is that it is a relic ionization signature from an AGN that has recently stopped accreting material and has ``turned off" \citep[e.g. ][]{Papaderos2013, Gomes2016, Schirmer2016, Keel2017, Ichikawa2019}.

\begin{table*}
\centering
\resizebox{\textwidth}{!}{%
\begin{tabular}{@{}rcccccccccccc@{}}
\toprule
\multicolumn{1}{c}{} & \multicolumn{3}{c}{\begin{tabular}[c]{@{}c@{}}RDAGN\\ ($\%$)\end{tabular}} & \multicolumn{3}{c}{\begin{tabular}[c]{@{}c@{}}Controls\\ ($\%$)\end{tabular}} & \multicolumn{3}{c}{\begin{tabular}[c]{@{}c@{}}LTGs\\ ($\%$)\end{tabular}} & \multicolumn{3}{c}{\begin{tabular}[c]{@{}c@{}}ETGs\\ ($\%$)\end{tabular}} \\ \midrule
\multicolumn{1}{r|}{R / R$_{e}$} & 0.2 & 0.6 & \multicolumn{1}{c|}{1.0} & 0.2 & 0.6 & \multicolumn{1}{c|}{1.0} & 0.2 & 0.6 & \multicolumn{1}{c|}{1.0} & 0.2 & 0.6 & \multicolumn{1}{c}{1.0} \\ \midrule
\multicolumn{1}{r|}{HII} & 0 & 0.33 & \multicolumn{1}{c|}{1.95} & 2.28 & 4.23 & \multicolumn{1}{c|}{7.82} & 0 & 0 & \multicolumn{1}{c|}{7.69} & 0 & 0.41 & \multicolumn{1}{c}{0.41} \\
\multicolumn{1}{r|}{Composite} & 5.86 & 7.82 & \multicolumn{1}{c|}{11.7} & 2.93 & 5.21 & \multicolumn{1}{c|}{4.56} & 7.69 & 20.0 & \multicolumn{1}{c|}{38.5} & 5.37 & 4.55 & \multicolumn{1}{c}{4.55} \\
\multicolumn{1}{r|}{LINER} & 6.51 & 2.93 & \multicolumn{1}{c|}{1.63} & 0 & 0 & \multicolumn{1}{c|}{0} & 15.4 & 4.62 & \multicolumn{1}{c|}{3.08} & 4.13 & 2.48 & \multicolumn{1}{c}{1.24} \\
\multicolumn{1}{r|}{Seyfert} & 0.98 & 0.98 & \multicolumn{1}{c|}{0.65} & 0 & 0 & \multicolumn{1}{c|}{0} & 3.08 & 3.08 & \multicolumn{1}{c|}{1.54} & 0.41 & 0.41 & \multicolumn{1}{c}{0.41} \\
\multicolumn{1}{r|}{LIER} & 85.7 & 87.3 & \multicolumn{1}{c|}{83.39} & 92.5 & 89.9 & \multicolumn{1}{c|}{87.0} & 69.2 & 72.3 & \multicolumn{1}{c|}{49.23} & 86.4 & 91.3 & \multicolumn{1}{c}{92.6} \\ \bottomrule
\end{tabular}
}
\caption{The percentages of galaxies with spaxels dominated by ionization classified as HII, Composite, LINER, Seyfert, and LIER for the RDAGN sample, the control sample, late-type AGN host galaxies, and early type AGN host galaxies at 0.2, 0.6 and 1.0 R$_{e}$. When the percentage equals 0, it indicates that the specific ionizing mechanism is not the dominant type at the given R$_{e}$.}
\label{tab: spaxel percentages}
\end{table*}

Galaxies dominated by composite spaxels (represented by the beige line in Figure \ref{fig:Excitation bar chart}) are the next most common type. In the RDAGN sample and the late-type RDAGN host galaxy sub-sample, the percentage of composite spaxel-dominated galaxies increases with increasing distance from the center of the galaxy. We find that the fraction of LTGs dominated by composite spaxels exhibits the largest increases in frequency with radius ($\sim$ 31$\%$ from 0.2 to 1.0 R$_{e}$). In the control sample, the percentage of galaxies dominated by composite-like excitation peaks at 0.6 R/R$_{e}$ (5.21$\%$). 

Compared to the entire control galaxies, we find that there are less RDAGN dominated by HII excitation (illustrated by the brown line in Figure \ref{fig:Excitation bar chart}). This could indicate that these RDAGN galaxies are more quenched than the control galaxies.

We find that only RDAGN exhibit LINER spaxels (light green colored line in Figure \ref{fig:Excitation bar chart}) and that the percentage of galaxies dominated by this excitation mechanism decreases with increasing R/R$_{e}$ ($\sim$ 6.5$\%$ to $\sim$ 1.6$\%$ from 0.2 to 1.0 R/R$_{e}$). 

Similar trends are observed for RDAGN galaxies dominated by Seyfert spaxels (mid-green line in Figure, although at smaller percentages than LINER spaxels (remains $<$ 1$\%$). It is not surprising that we do not find any Seyfert or LINER dominated control galaxies because our selection excluded galaxies dominated by LINER and Seyfert excitation in the central 3$^{\prime\prime}$ of the SDSS fibre (see Section \ref{subsec:Control sample criteria}).

\subsection{SFR surface density ($\sum$SFR)}
\label{subsec:SFR surface denisty}

In order to obtain the SFR surface density $\sum$SFR), we calculated the SFR in each spaxel using the extinction-corrected L$_{\mathrm{H\alpha}}$ equation from \citet{Kennicutt1998}:
\begin{equation}
    \sum \mathrm{SFR} = 7.9 \times 10^{-42} \times L(\mathrm{H}\alpha),
\label{eq: Kennitcutt SFR}
\end{equation}
where L(H$\alpha$) is in units of erg s$^{-1}$. We correct H$\alpha$ emission for extinction ($\lambda$ = 6563 \text{\r{A}}) in magnitudes calculated by \citet{Cardelli1989}:
\begin{equation}
    A_{\lambda} = A_{V}\Big(a+\frac{b}{2.87}\Big)
\label{eq: Alam}
\end{equation}
where A$_{V}$ is derived by comparing the ratio of extinction for the observed fluxes of H$\alpha$ (F(H$\alpha$)) and H$\beta$ (F(H$\beta$)) to theoretical intrinsic value from case B recombination of \citet{Osterbrock2006}:
\begin{equation}
    A_{V}= 7.23 \times \log\Big[ \frac{F(H\alpha)}{F(H\beta)} \times\frac{1}{2.87}\Big].
\label{eq: Av} 
\end{equation}

From there, we calculated the extinction-corrected F(H$\alpha$) (F(H$\alpha$)$_{0}$; in units 10$^{-17}$ erg s$^{-1}$ cm$^{-2}$ spaxel$^{-1}$):
\begin{equation}
    F(H\alpha)_{0} = F(H\alpha)\times(10^{0.4A_{\lambda}})
\label{eq: E corrected F} 
\end{equation}
and finally the extinction-corrected L(H$\alpha$):
\begin{equation}
    L(H\alpha) = 1\times 10^{-17} \times F(H\alpha)_{0}\times 4\pi d_{cm}^{2}
\label{eq: E corrected L} 
\end{equation}
where d$_{cm}$ is the luminosity distance in centimeters at the redshift of each galaxy, calculated using \verb|Astropy|’s\footnote{Publicly available software package for the Python programming language: \url{https://www.astropy.org/}} function \verb|cosmo.luminosity_distance()|.
To convert the angular size of each spaxel to physical size, we calculated the following scale-factor using the small angle approximation and the galaxy’s luminosity distance in kpc, d$_{kpc}$. The area of the spaxel was then determined by multiplying the scaling relation by the angular size of the spaxel (0.5$^{\prime\prime}$ for MaNGA IFU) squared. Finally, after calculating the SFR in HII and composite spaxels, we divided each spaxel by its physical size to obtain the $\sum$SFR.

In the bottom panels of Figure \ref{fig:spatial maps}, we present the surface distribution of the $\sum$SFR. Unlike the spatial maps for the late-type RDAGN example and its assigned control galaxy, the maps for the early-type RDAGN host and control are blank. This is expected because neither the early-type RDAGN host galaxy nor its control contained HII or composite spaxels.

\begin{figure*}[!ht]
    \centering
    \includegraphics[width = \textwidth,  ]{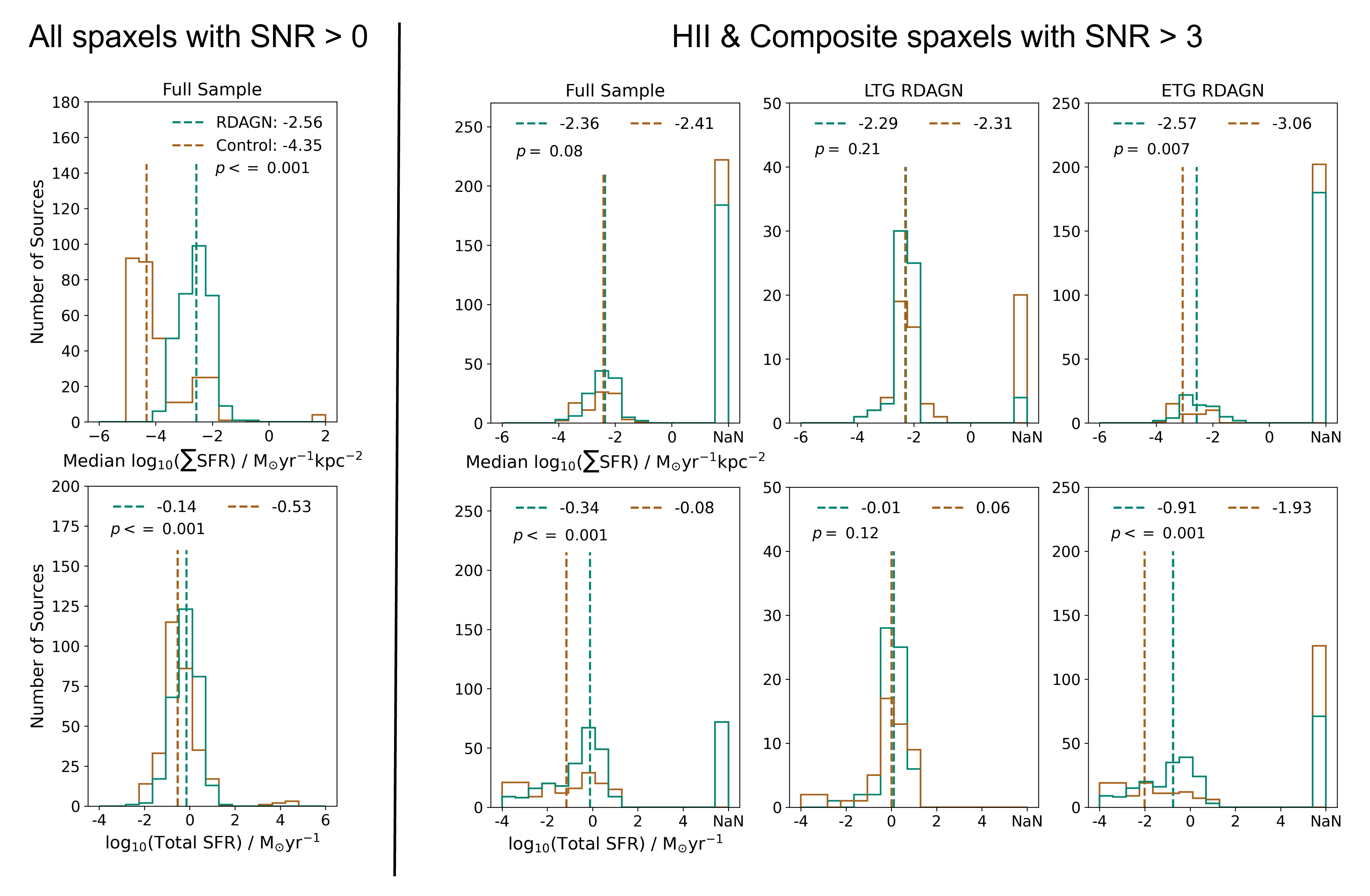}
    \caption[$\sum$SFR histograms, Total SFR, and fractional difference in total SFR between the RDAGN and control sample.]{\textit{Left of the black vertical line: } Distribution of the median $\sum$SFR (top panel) and for the total SFR (bottom panel) for RDAGN galaxies (green) and their controls (brown). We calculate these quantities using all spaxels with a S/N > 0 in all emission-lines used for classification. \textit{Right of the black vertical line from left to right: } Distribution of the median $\sum$SFR (top panel) and for the the total SFR (bottom panel) for RDAGN galaxies (green) and their controls (brown) for the entire sample, LTGs, and ETGs. These quantities are derived from spaxels with a S/N > 3. When no spaxels in the galaxy meet the relevant criteria, the median or total SFR is set to ``NaN''. }
    
    \label{fig:ESFR Histograms}
\end{figure*}

In Figure \ref{fig:ESFR Histograms}, we present histograms for the median $\sum$SFR and total SFR (sum of SFR across every spaxel) for the RDAGN sample and control sample and for samples subdivided according to morphology. To the left of the black vertical line, these quantities were derived using all spaxels with a S/N $>$ 0, and to the right of the line, only HII and composite spaxels with a S/N $>$ 3. We show the results from these different scenarios to gauge whether or not our choice to measure the SFR spaxels with HII and composite spaxels affected our final result. We report a statistically significant difference between the median $\sum$SFR of the RDAGN and the control sample when all spaxels are considered, but no difference in the total SFRs. The RDAGN show higher median $\sum$SFR than the control sample ($\sum$SFR =$10^{-2.56}$ compared to $10^{-4.35}$ M$_{\odot}$yr$^{-1}$kpc$^{-2}$, which could indicate either that there are regions with enhanced SFR within our RDAGN (signs of positive feedback), or, that calculating SFR from H$\alpha$ in these cases is not reliable. We interpret these results as confirmation of our choice to measure the SFR in HII and composite spaxels.
We chose not to show the RLAGN sub-sample on these panels because there are too few galaxies in the sample for any differences in the $\sum$SFR and total SFR between the AGN host galaxies and the control galaxies to be called statistically significant. 

When considering HII and Composite spaxels with S/N$>$3, we find that the average $\sum$SFR for RDAGN galaxies is -2.36 in logarithmic units of M$_{\odot}$yr$^{-1}$kpc$^{-2}$, which is higher than the controls' value of -2.41. We find that the total SFR for RDAGN ranges between $\sim$ 10$^{-4.23}$ M$_{\odot}$yr$^{-1}$ and 10$^{1.09}$ M$_{\odot}$yr$^{-1}$. The total SFR of the controls range from $\sim$ 10$^{-5.25}$ to 10$^{1.21}$ M$_{\odot}$yr$^{-1}$. 

Towards assessing the probability that the RDAGN sample and the control sample were drawn from the same parent population, we performed a two-sample Anderson-Darling (A-D) test. When the A-D statistic is less than the critical value at the specified significance level, the null hypothesis--that the $\sum$SFR RDAGN sample and the control sample were drawn from the same distribution-- cannot be rejected in favor of the alternative hypothesis, which is that the distributions of the two samples are different. Before performing the test, we set the reference significance level to 0.05. For the $\sum$SFR of entire RDAGN and control samples, which is presented in the top panel of Figure \ref{fig:ESFR Histograms}, the A-D statistic is $\sim$ 0.08, which is less than the critical value at $p =$ 0.05 ($\sim$4.59). Therefore, the null hypothesis is not rejected and we concluded that the distributions of the $\sum$SFR for the RDAGN and the control galaxies are statistically similar. We found the same conclusions for the late-type RDAGN sub-sample and their control galaxies. Conversely, we found that the distribution of $\sum$SFR are statistically different for the early type sub-sample of RDAGN and their controls (the null hypothesis can be rejected at the $>5$\% level). The early type RDAGN galaxies tend to have higher $\sum$SFR values (median value of -2.57 in logarithmic units of M$_{\odot}$yr$^{-1}$kpc$^{-2}$) than the $\sum$SFR of their assigned control galaxies, which averages at -3.06 in logarithmic units of M$_{\odot}$yr$^{-1}$kpc$^{-2}$. 

For the distribution of total SFRs, which are shown in the bottom panels of Figure \ref{fig:ESFR Histograms}, only the late-type sub-sample of RDAGN and their control galaxies exhibit a statistically similar distribution based on the A-D test (p $\sim$ 0.12). While the distributions for the entire RDAGN, the early-type RDAGN sub-sample, and their assigned control galaxies most likely reveal physical differences, our analyses would benefit from more accurate SFR measurements, which would require decomposing each spectrum into SF, AGN, and shock components. 

Our results are both consistent and at variance with the findings of \citet{doNascimento2019}, which use the MaNGA AGN and control sample selected by \citet{Rembold2017}. By interpreting the p-values of A-D tests, both this study and \citet{doNascimento2019} find that the $\sum$SFR are statistically similar for the AGN and controls. We report, however, a wider range of total SFRs; \citet{doNascimento2019} find both the AGN and control sample to range in SFR from 10$^{-3}$ to 10$^{1}$ M$_{\odot}$yr$^{-1}$.

Neither our study nor that of \citet{doNascimento2019} accounted for disk inclination when calculating $\sum$SFRs, which could cause SFRs to be underestimated by a factor of $\sim$ 0.2-0.4 dex due to not completely correcting dust attenuation \citep[e.g. ][]{Morselli2016}. However, given that the inclination of the RDAGN and of their assigned control sample were visually matched, our comparison does not suffer from a large inclination bias. Furthermore, both this study and \citet{doNascimento2019} only consider HII and composite spaxels when calculating $\sum$SFR. The composite spaxels could be contaminated by shocks. Following \citet{Davies2017}, future work could include calculating a more accurate SFR by decomposing the nuclear spectra into SF, AGN, and shock components.

\section{Stellar age gradient}
\label{section:Stellar age gradient}

\begin{figure}[!t]
    \centering
    \includegraphics[width = 0.48\textwidth, trim= 0cm 0cm 0cm 0cm,clip]{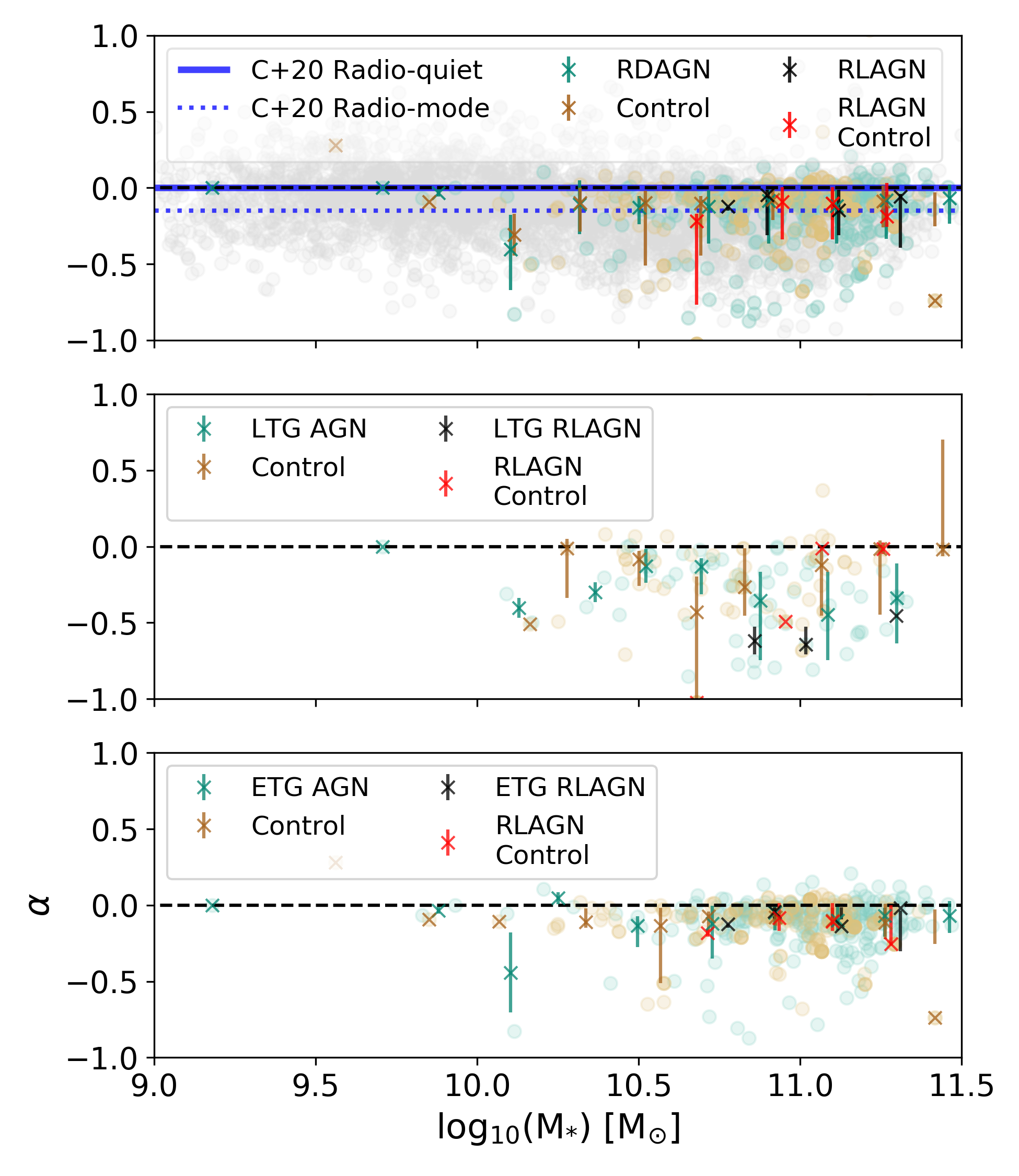}
    \caption[Stellar age gradients ($\alpha$) from Pipe3D in M$_{*}$ bins of 0.2 dex.]{Stellar, light-weighted age gradient ($\alpha$) in M$_{*}$ bins of 0.2 dex for RDAGN host galaxies (green), the full control sample (brown), RLAGN galaxies (black), RLAGN control galaxies (red), and the entire Pipe3D VAC (grey). A horizontal line is plotted at $\alpha$ = 0 for reference. The x's represent the median value in each bin and the error bars represent the standard deviation of the sample. We provide the median $\alpha$ values for radio-quiet ($\alpha \sim$ 0) and radio-mode ($\alpha \sim$ -0.15) AGN from \citet{Comerford2020} (C+20) in the solid and dotted blue lines, respectively. }
    \label{fig:age gradient Full}
 \end{figure}   
 
 To find evidence for suppressed SF in RDAGN host galaxies and potentially in the control galaxies, we examined how the age of the stellar populations change as a function of galactocentric distance. For this analysis, we use the gradient of the light-weighted log-age of the stellar population within a galactocentric distance of 0.5-2.0 R$_{e}$ (hereafter $\alpha$) from the Pipe3D VAC.  When $\alpha$ is negative, the stellar populations become younger with distance from the center of the galaxy. Conversely, a positive age gradient indicates the stellar populations become older with increasing distance away from the galaxy's center. We compare $\alpha$ in stellar mass bins of 0.2 dex because previous studies have demonstrated that a correlation exists between a galaxy's M$_{*}$ and stellar age gradient \citep[e.g. ][]{Gonzalez2014, Zheng2017, Goddard2017} and for comparison purposes with the Pipe3D stellar, light-weighted age gradients binned in M$_{*}$ for radio-quiet and radio-mode AGN host galaxies from \citet{Comerford2020}. In Figure \ref{fig:age gradient Full}, we present the Pipe3D stellar, light-weighted age gradient in M$_{*}$ bins of 0.2 dex for the entire RDAGN and control samples as well as for RDAGN sub-samples and their controls.

 We find that the average stellar age gradients for the RDAGN sample and control sample as measured by Pipe3D are negative. Their average values ($\alpha \sim$ -0.101 for RDAGN and $\alpha \sim$ -0.097 for the controls) are nearly identical (see Table \ref{tab:Pipe3D Results} for all averages), indicating that the stellar populations within the RDAGN sample and the control sample become younger with distance from the center. These results may point to the inside-out suppression SF in these galaxies. Moreover, the consistency between the age gradient values between the AGN and controls indicates that there is no clear correlation between the current AGN activity and their host galaxies' star formation history.

The average $\alpha$ value for late-type AGN host galaxies ($\alpha \sim$ -0.294) is significantly steeper than early type AGN host galaxies ($\alpha \sim$ -0.070), which agrees with the results from previous MaNGA investigations \citep{Goddard2017, Parikh2021}. The negative radial stellar age gradients in LTGs are consistent with inside-out growth of the disk \citep{Gonzalez2015}. On the other hand, strong AGN feedback can stop star formation in the galaxy's centre, and this inside-out quenching may also result in a negative age gradient \citep{Comerford2020}.
 
 The number of fibers in an IFU bundle affects the accuracy of the estimate of $\alpha$ \citep{IbarraMedel2019}. So, IFUs with a larger fiber bundle will have a more accurate measurement of $\alpha$. \citet{Comerford2020} have investigated the magnitude of this effect on their sample of 406 MaNGA-AGN by looking at the stellar age gradients of galaxies that were observed with the largest MaNGA fiber bundle size (127 fibers, commensurate with a diameter of 32$^{\prime\prime}$.5). \citet{Comerford2020} found that $\alpha$ decreased by $\sim$ 0.05, but that did not change their result that radio-mode AGN host galaxies have more negative stellar age gradients when compared to radio-quiet AGN host galaxies. We find that the age gradients of RDAGN and control galaxies observed with the largest MaNGA fiber bundle decrease by $\sim$ 0.10. These RDAGN and control galaxies have an identical average age gradient of $\alpha \sim$ -0.20 $\pm$ 0.30. By checking the magnitude of the effect of IFU fiber bundle size, we have reconfirmed the striking similarity between the RDAGN sample and control sample.

Residual AGN contamination can bias the stellar population fits \citep[e.g.][]{Cardoso2017}. However, quantifying and rectifying this bias is beyond the scope of this work.

 \begin{table}[!t]
\centering
\resizebox{\columnwidth}{!}{%
\begin{tabular}{@{}lcc@{}}
\toprule
Sample & Median $\Delta$ log$_{10}$(SFR) & Median $\alpha$ \\ \midrule
RDAGN & -1.51 $\pm$ 3.20 & -0.097 $\pm$ 0.226 \\
Full Control & -2.29 $\pm$ 3.03 & -0.100 $\pm$ 0.221 \\
RLAGN & -2.35 $\pm$ 3.15 & -0.09 $\pm$ 0.211 \\
RLAGN Controls & -2.38 $\pm$ 1.75 & -0.11 $\pm$ 0.186 \\
D$_{n}$4000 vs. L$_{1.4\;\mathrm{GHz}}$ / M$_{*}$ AGN & -1.52 $\pm$ 3.31 & -0.093 $\pm$ 0.217 \\
{[}NII{]} BPT AGN & -0.359 $\pm$ 3.37 & -0.128 $\pm$ 0.211 \\
L$_{\mathrm{H\alpha}}$ vs. L$_{150\;\mathrm{MHz}}$ AGN & -2.35 $\pm$ 3.15 & -0.092 $\pm$ 0.211 \\
WISE Color-Color AGN & -1.51 $\pm$ 3.22 & -0.094 $\pm$ 0.224 \\
LTG AGN & -0.465 $\pm$ 3.47 & -0.263 $\pm$ 0.238 \\
ETG AGN & -1.74 $\pm$ 3.09 & -0.075 $\pm$ 0.201 \\ \bottomrule
\end{tabular}%
}
\caption[Median distance from the star-forming main-sequence.]{Median distance from the star-forming main-sequence, where $\Delta$ log$_{10}$(SFR) = log$_{10}$(SFR$_{sample}$) $-$  log$_{10}$(SFR$_{SFMS}$) and for the average stellar, light-weighted age gradient ($\alpha$) values. All errors represent the standard deviation of the sample.}
\label{tab:Pipe3D Results}
\end{table}

\section{Discussion}
\label{section:Discussion}

In establishing whether or not AGN are responsible for quenching massive galaxies, we compare the SF properties of radio-detected AGN with non-active galaxies of similar stellar mass, redshift, visual morphology, and inclination. AGN remain a key ingredient in cosmological models of galaxy evolution to reproduce the observed stellar mass and luminosity function and to prevent the formation of over-massive galaxies. However, the observational perspective has yielded mixed results, and therefore, the consensus on the effect of an AGN on their host galaxies' SFR has yet to be agreed upon. One of the most interesting results of our paper is that both radio-detected AGN and control galaxies typically lie below the main-sequence, have broad SFR distributions, and exhibit negative stellar, light-weighted age gradients. 

One possible explanation for the statistical similarity between the quenching patterns of our AGN-host galaxies and the control sample of non-active galaxies is the visibility timescales of AGN feedback. Much remains unclear about the timescales of the duty cycle of AGN, the duration of visible AGN episodes, the spatial scale at which these interactions occur and AGN variability \citep[e.g, ][]{AlexanderHickox2012, Hickox2014,Sartori2018}. Studies \citep[e.g. ][ and references therein]{Sanchez2018, Lacerda2020} suggest that the timescales required to quench SF and the triggering of AGN activity could be completely different. Moreover, the fact that RLAGN appear to preferentially reside in ETGs, and that they are considerably more quenched than just RDAGN might suggest that radio activity is supported for a longer period, and quenching has occurred earlier in their host galaxies' lifetime. Additionally, how long it takes for AGN to have an observed effect on SF is still an unanswered question. Hence, the timescale of the suppression of SF from an AGN episode -- or multiple AGN episodes -- might be longer than the timescale of observable AGN activity \citep{Harrison2017}.  Furthermore, the flickering on and off of AGN may also play a role in maintaining galaxy quiescence, which could explain why we see little differences in the AGN and control galaxies.

An abundance of physical mechanisms have been evoked to explain galaxy quiescence. In our study, we do not expect that environmental effects play a significant role in quenching our RDAGN and control galaxies given their average stellar masses and redshifts \citep[$\sim$ 10$^{11}$ M$_{*}$ and \textit{z} $\sim$ 0, respectively; ][]{Peng2010}. Furthermore, results from SDSS-IV MaNGA-DR15 and the GASP survey suggest that for environmental quenching, quenching is expected to occur from the outside-in \citep[e.g. ][]{Bluck2020a, Vulcani2020}. Recent studies \citep[e.g.][]{Bluck2018, Bluck2020a, Bluck2020b} have demonstrated that there is indeed a connection between quenching and the presence of central supermassive black hole, which is consistent with expected models of quenching via AGN feedback. Our study reveals a similarity in the star-forming properties of radio-detected AGN host galaxies and non-active control galaxies, which may indicate that AGN feedback is likely not the only origin of inside-out quenching. Additionally, our results suggest that the effect of mass quenching from negative AGN feedback is indistinguishable from the effect of other mass quenching mechanisms such as virial shock heating in massive dark matter haloes, which prevents the accretion of cold gas onto galaxies \citep[e.g. ][]{Birnboim2003, Keres2005, Dekel2006, Birnboim2007, Dekel2008, Keres2009}. Alternatively, SF may be quenched in galaxies without the expulsion and/or heating of gas. Instead, SF can be halted as a galaxy transitions to being dominated by a stellar spheroid, which stabilizes the gas disk and prevents it from fragmenting into star-forming clumps \citep[i.e. morphological quenching; ][]{Martig2009}.

Finding direct evidence for AGN feedback quenching SF in local radio galaxies would naturally be difficult because they predominately reside in massive galaxies where star formation has already been quenched. Additionally, the bulk of the energetic impact of a radio AGN is injected into the hot phase of their host galaxies' halo, where it only has a long-term effect on the SF history of the host galaxy. 

\subsection{Comparison to other IFS investigations of AGN}

Throughout this work, we compare our sample of RDAGN host galaxies to existing MaNGA AGN Catalogs (see Section \ref{subsec:Global properties}). Several of these MaNGA AGN catalogs \citep[][]{Rembold2017, Wylezalek2018, Sanchez2018}  select AGN with optical emission line ratios and cuts in the EW(H$\alpha$), and \citet{Comerford2020} take a multi-wavelength approach. The main differences we see among these studies and our own is that the selection method determines the number of sources that are considered AGN host galaxies, and the intrinsic global properties they select for. 

Our results are both consistent and in disagreement with those presented in \citet{doNascimento2019}, which compare the optically-selected AGN sample from \citet{Rembold2017} with a control sample of non-active galaxies with similar global properties as each AGN host galaxy. Similar to our results, \citet{doNascimento2019} find no differences in SFR between optically-selected, late-type AGN host galaxies and their controls. However, \citet{doNascimento2019} report that early-type AGN host galaxies typically exhibit higher SFRs and larger ionized gas masses than their assigned control galaxies. They attribute this result to AGN and SF activity being fueled by the same reservoir of gas. Hence, \citet{doNascimento2019} suggest that it is unlikely that negative AGN feedback is occurring in the \citet{Rembold2017} MaNGA AGN sample. While our results do not indicate that AGN selected based on
their current activity are responsible for suppressing their host galaxies’ star formation, they support the maintenance mode role that RDAGN are expected to play in the local Universe. We believe the difference in our findings for early-type AGN host galaxies is a result of sample selection methods.

We find that RDAGN, and classical RLAGN preferentially reside in ETGs, lie below the SFMS, and exhibit younger stellar populations with increasing distance from the host galaxies' centers. Our work compliments the findings presented in \citet{Comerford2020}, which compare the SF properties of radio-mode and radio-quiet AGN host galaxies. They find that radio-quiet and radio-mode AGN preferentially reside in LTGs and ETGs, respectively, both populations fall below the SFMS, although radio-mode AGN host galaxies lie further below the SFMS, and that radio-mode AGN exhibit older stellar populations and have more negative stellar age gradients than the radio-quiet sample. From these results, \citet{Comerford2020} suggest that radio-mode AGN played a role in quenching star formation in their host galaxies' pasts. Despite showing similar, albeit less obvious signs of past quenching, \citet{Comerford2020} do not provide a suggestion for the role radio-quiet AGN played in their host galaxies' past. Our study is different in that we compared these radio-selected AGN to \textit{non-active} galaxies that match the stellar mass, redshift, visual morphology, and inclination of their RDAGN counterpart. Furthermore, our comparison to \textit{non-active} galaxies, and our finding that there is no statistically significant difference between these two populations, is a more robust evaluation of the role RDAGN played in the star formation quenching in the past.

\citet{Sanchez2018}, and other IFS investigation of the role of AGN feedback in quenching SF \citep[e.g. ][and references therein]{Lacerda2020} have found that we cannot yet establish a causal connection between the presence of an AGN and the quenching of their host galaxies' SF. Instead, AGN activity and SF processes present an apparent co-evolution, which could be affected by the growth of galactic bulges. Similarly, the results presented here do not establish a casual connection between AGN activity and the halting of SF. Ours points to a scenario where there could be multiple quenching mechanisms occurring simultaneously, and where AGN play a role maintaining quiescence.

\section{Conclusions}
\label{section:Conlcusions}

In this work, we have investigated whether negative AGN feedback is responsible for quenching massive galaxies. We combined the LoTSS DR2 and MaNGA DR16 data to form a sample of 1250 galaxies from which 307 RDAGN host galaxies were identified by combining selection techniques using global emission-line properties, radio luminosities, and WISE mid-infrared luminosities. Our investigation is the largest, IFS multi-wavelength study of AGN that has a control sample of non-active galaxies. Furthermore, thanks to the low frequencies and sensitivities reached by LOFAR, this study detects fainter radio emission from lower-powered jets -- as well as remnant emission from sources that have recently shut-off their jet activity -- than what was previously possible for radio surveys (e.g. NVSS, FIRST, etc.). Therefore, this work has resulted in significant progress towards understanding the effect of AGN feedback in a representative sample of low-luminosity AGN host galaxies.

We spatially mapped the dominant excitation mechanism of emission-line gas in RDAGN and control galaxies by combining the [NII] BPT, [SII] BPT, and the WHAN diagram. In regions ionized by star-formation, we calculated the SFR surface density ($\sum$SFR) using the dust corrected luminosity of H$\alpha$. We also used cumulative and gradient properties taken from the Pipe3D value added catalog to determine the relation of these galaxies to the star-forming main-sequence and how the age of their stellar populations changes as a function of galactocentric radius. Our main results are summarized below:

\begin{enumerate}
    
    \item RDAGN and control galaxies display a statistically similar distribution for the median star-formation rate surface density ($\sum$SFR). The fractional difference in $\sum$SFR of the RDAGN and its assigned control galaxy reveal that RDAGN host galaxies typically have higher SFRs. 
    
    \item RDAGN host galaxies lie below the star-forming main-sequence, which suggests that RDAGN occupy galaxies with suppressed star-formation. RDAGN host galaxies have an average $\Delta$ log$_{10}$(SFR) $\sim$ -1.5, while control galaxies fall further below the star-forming main-sequence at an average $\Delta$ log$_{10}$(SFR) $\sim$ -2.3.

    \item The average SFR for RDAGN, as measured by Pipe3D, is higher ($\sim$ 10$^{-1}$ M$_{\odot}$ yr$^{-1}$) than the average SFR for the control sample of non-active galaxies ($\sim$ 10$^{-1.8}$ M$_{\odot}$ yr$^{-1}$). Taken together with the preceding points, we find no direct evidence that SF is quenched in RDAGN host galaxies. In fact, when compared to the control galaxies, our results may point to either the effect of negative AGN feedback has not yet fully halted SF or positive AGN feedback might be occurring in some late-type systems.
    
    \item The average stellar, light-weighted age gradient for the RDAGN and control galaxies are identical at $\alpha \sim -$ 0.10. The negative age gradient implies that the stellar populations in the centers of galaxies are older than the populations on the outskirts. These results may point to inside-out quenching of star formation in both samples. We find that early type RDAGN host galaxies have a relatively flat average age gradient ($\alpha \sim$ -0.08) whereas LTGs exhibit a steeper gradient ($\alpha \sim$ -0.26).

\end{enumerate}

This work demonstrates that the physical mechanisms behind the origin of the quenching of SF are yet to be fully understood. To further our understanding of how these RDAGN and their host galaxies are co-evolving, a detailed kinematic analysis could help determine the prevalence and velocity of outflows. Furthermore, the RDAGN sample in this work includes galaxies that have both AGN activity and some star formation activity. Additional work is needed to decompose the radio emission into that coming from SF and that from jets. This will involve using LOFAR's international baselines to obtain high (subarcsecond) resolution images, which will allow us to identify genuine AGN emission and its effect on its host galaxy. We have already begun additional investigations on the molecular gas content of a sub-sample of these RDAGN host galaxies (Leslie et al. in prep.). We intend to use these observations to determine whether there is a deficiency of molecular gas in the central regions of RDAGN galaxies, which would quench central SF. Additionally, we could establish whether radio-mode AGN suppress star formation either through their jet's mechanical energy heating the surrounding ISM preventing molecular gas from radiatively cooling or if AGN-driven outflows expel the molecular gas out of the galaxy by correlating radio source size with stellar age and determining the SF efficiency.  

\section{Acknowledgements}
\label{section:acknowledgements}

We thank the referee for valuable comments on the
paper. CRM thanks Dr. Dominika Wylezalek for providing the IDs for the \citet{Wylezalek2018} sample. IP acknowledges support from INAF under the SKA/CTA PRIN “FORECaST” and the PRIN MAIN STREAM “SAuROS” projects. MJH acknowledges support from STFC [ST/V000624/1]. KM has been supported by the National Science Centre (UMO-2018/30/E/ST9/00082). MB acknowledges support from INAF under the SKA/CTA PRIN “FORECaST” and the PRIN MAIN STREAM “SAuROS” projects and from the Ministero degli Affari Esteri e della Cooperazione Internazionale - Direzione Generale per la Promozione del Sistema Paese Progetto di Grande Rilevanza ZA18GR02.

LOFAR (van Haarlem et al. 2013) is the Low Frequency Array designed and constructed by ASTRON. It has observing, data processing, and data storage facilities in several countries, which are owned by various parties (each with their own funding sources), and that are collectively operated by the ILT foundation under a joint scientific policy. The ILT resources have benefited from the following recent major funding sources: CNRS-INSU, Observatoire de Paris and Université d'Orléans, France; BMBF, MIWF-NRW, MPG, Germany; Science Foundation Ireland (SFI), Department of Business, Enterprise and Innovation (DBEI), Ireland; NWO, The Netherlands; The Science and Technology Facilities Council, UK; Ministry of Science and Higher Education, Poland; The Istituto Nazionale di Astrofisica (INAF), Italy.
This research made use of the Dutch national e-infrastructure with support of the SURF Cooperative (e-infra 180169) and the LOFAR e-infra group. The Jülich LOFAR Long Term Archive and the German LOFAR network are both coordinated and operated by the Jülich Supercomputing Centre (JSC), and computing resources on the supercomputer JUWELS at JSC were provided by the Gauss Centre for Supercomputing e.V. (grant CHTB00) through the John von Neumann Institute for Computing (NIC). 
This research made use of the University of Hertfordshire high-performance computing facility and the LOFAR-UK computing facility located at the University of Hertfordshire and supported by STFC [ST/P000096/1], and of the Italian LOFAR IT computing infrastructure supported and operated by INAF, and by the Physics Department of Turin university (under an agreement with Consorzio Interuniversitario per la Fisica Spaziale) at the C3S Supercomputing Centre, Italy.

Funding for the Sloan Digital Sky 
Survey IV has been provided by the 
Alfred P. Sloan Foundation, the U.S. 
Department of Energy Office of 
Science, and the Participating 
Institutions. SDSS-IV acknowledges support and 
resources from the Center for High 
Performance Computing  at the 
University of Utah. The SDSS 
website is www.sdss.org.

SDSS-IV is managed by the 
Astrophysical Research Consortium 
for the Participating Institutions 
of the SDSS Collaboration including 
the Brazilian Participation Group, 
the Carnegie Institution for Science, 
Carnegie Mellon University, Center for 
Astrophysics | Harvard \& 
Smithsonian, the Chilean Participation 
Group, the French Participation Group, 
Instituto de Astrof\'isica de 
Canarias, The Johns Hopkins 
University, Kavli Institute for the 
Physics and Mathematics of the 
Universe (IPMU) / University of 
Tokyo, the Korean Participation Group, 
Lawrence Berkeley National Laboratory, 
Leibniz Institut f\"ur Astrophysik 
Potsdam (AIP),  Max-Planck-Institut 
f\"ur Astronomie (MPIA Heidelberg), 
Max-Planck-Institut f\"ur 
Astrophysik (MPA Garching), 
Max-Planck-Institut f\"ur 
Extraterrestrische Physik (MPE), 
National Astronomical Observatories of 
China, New Mexico State University, 
New York University, University of 
Notre Dame, Observat\'ario 
Nacional / MCTI, The Ohio State 
University, Pennsylvania State 
University, Shanghai 
Astronomical Observatory, United 
Kingdom Participation Group, 
Universidad Nacional Aut\'onoma 
de M\'exico, University of Arizona, 
University of Colorado Boulder, 
University of Oxford, University of 
Portsmouth, University of Utah, 
University of Virginia, University 
of Washington, University of 
Wisconsin, Vanderbilt University, 
and Yale University. This project makes use of the MaNGA-Pipe3D dataproducts. We thank the IA-UNAM MaNGA team for creating this catalogue, and the ConaCyt-180125 project for supporting them. 


\bibliographystyle{aa}
\bibliography{RDAGN.bib}

\clearpage
\begin{appendix}
\label{A:appendix}
 
\section{Spatially resolved maps and diagrams for all 307 RDAGN galaxies}
\label{A:EL_examples}

Resolved ionization classification maps and $\sum$SFR maps for the full sample of 307 RDAGN galaxies and their assigned control galaxies are available upon request from the authors.

\section*{Extra Tables}
\onecolumn
\begin{longtable}{@{}|cc|cc|cc|cc|@{}}
\toprule
\begin{tabular}[c]{@{}c@{}}RLAGN\\ plateifu\end{tabular} & \begin{tabular}[c]{@{}c@{}}Control\\ plateifu\end{tabular} & \begin{tabular}[c]{@{}c@{}}RLAGN\\ plateifu\end{tabular} & \begin{tabular}[c]{@{}c@{}}Control\\ plateifu\end{tabular} & \begin{tabular}[c]{@{}c@{}}RLAGN\\ plateifu\end{tabular} & \begin{tabular}[c]{@{}c@{}}Control\\ plateifu\end{tabular} & \begin{tabular}[c]{@{}c@{}}RLAGN\\ plateifu\end{tabular} & \begin{tabular}[c]{@{}c@{}}Control\\ plateifu\end{tabular} \\* \midrule
\endfirsthead
\multicolumn{8}{c}%
{{\bfseries Table \thetable\ continued from previous page}} \\
\toprule
\begin{tabular}[c]{@{}c@{}}RLAGN\\ plateifu\end{tabular} & \begin{tabular}[c]{@{}c@{}}Control\\ plateifu\end{tabular} & \begin{tabular}[c]{@{}c@{}}RLAGN\\ plateifu\end{tabular} & \begin{tabular}[c]{@{}c@{}}Control\\ plateifu\end{tabular} & \begin{tabular}[c]{@{}c@{}}RLAGN\\ plateifu\end{tabular} & \begin{tabular}[c]{@{}c@{}}Control\\ plateifu\end{tabular} & \begin{tabular}[c]{@{}c@{}}RLAGN\\ plateifu\end{tabular} & \begin{tabular}[c]{@{}c@{}}Control\\ plateifu\end{tabular} \\* \midrule
\endhead
\bottomrule
\endfoot
\endlastfoot
9891-3704 & 9507-6103 & 9026-12705 & 8439-12702 & 8716-3702 & 8721-6102 & 8447-1902 & 8440-3702 \\
9891-3702 & 8713-3702 & 9025-9101 & 9045-12705 & 8712-12705 & 8978-12704 & 8447-12704 & 8315-12702 \\
9883-9101 & 9044-12703 & 9025-12704 & 8984-12705 & 8712-12704 & 8602-12705 & 8446-3701 & 8718-3702 \\
9883-6101 & 8326-3702 & 9025-12701 & 9501-12703 & 8711-12704 & 8243-12703 & 8445-12702 & 8262-6103 \\
9883-3702 & 8440-3702 & 9024-3703 & 8713-3702 & 8613-6102 & 8943-3701 & 8439-3704 & 8568-9102 \\
9881-9101 & 8149-12704 & 9024-12702 & 9085-6104 & 8613-3702 & 8944-3702 & 8341-12702 & 8451-6102 \\
9881-3701 & 8588-3701 & 9002-9101 & 9038-12703 & 8606-9101 & 8555-6101 & 8335-9101 & 8315-12702 \\
9871-12702 & 8602-12705 & 9002-3703 & 8989-6104 & 8604-6102 & 8997-6101 & 8335-6103 & 8315-12702 \\
9870-1901 & 8547-1902 & 9002-12703 & 8613-12704 & 8604-12703 & 8947-12704 & 8333-9102 & 8141-6104 \\
9868-6104 & 9041-3701 & 9002-12702 & 8315-12702 & 8602-12701 & 8602-12705 & 8333-6103 & 8326-9101 \\
9868-3704 & 8485-3703 & 9000-9102 & 8613-12704 & 8601-12704 & 8464-9101 & 8333-6101 & 8326-3702 \\
9865-9101 & 9487-12705 & 9000-12703 & 8139-9102 & 8600-12703 & 8936-12702 & 8333-3703 & 9185-3703 \\
9864-6104 & 8258-3703 & 9000-12701 & 8548-3701 & 8597-9101 & 8452-6103 & 8333-12704 & 8948-6104 \\
9864-3702 & 8313-3702 & 8999-3702 & 9891-3701 & 8597-3704 & 8566-6104 & 8333-12701 & 8715-12703 \\
9864-12705 & 8332-6101 & 8997-9101 & 8139-9102 & 8597-3703 & 9182-6103 & 8332-6104 & 8455-6104 \\
9864-12702 & 8452-6103 & 8997-6104 & 8612-6101 & 8597-3701 & 9486-1902 & 8332-12705 & 8312-9101 \\
9864-12701 & 8939-6103 & 8997-6103 & 8465-3703 & 8597-12702 & 9505-12701 & 8331-9101 & 8315-12702 \\
9510-6104 & 9507-6103 & 8997-6102 & 8989-6104 & 8595-6101 & 9041-3701 & 8331-3702 & 8258-3703 \\
9510-12705 & 8131-6103 & 8997-1902 & 9501-3702 & 8595-12704 & 9184-9101 & 8331-3701 & 9182-1901 \\
9508-9102 & 8996-3702 & 8995-3703 & 8715-3703 & 8592-12703 & 8330-12705 & 8331-12701 & 9026-12704 \\
9508-12702 & 9026-6104 & 8995-12705 & 8443-12703 & 8591-6101 & 8252-3702 & 8330-6103 & 9509-6102 \\
9507-3701 & 8980-6102 & 8995-12703 & 8315-12702 & 8591-3701 & 8938-6101 & 8325-6101 & 8445-6102 \\
9485-6103 & 9028-6104 & 8993-12705 & 8249-12703 & 8588-6104 & 8465-3703 & 8323-6101 & 8947-12704 \\
9485-6102 & 9041-3701 & 8992-9102 & 8149-12704 & 8588-6102 & 8313-3702 & 8323-1902 & 9029-6101 \\
9485-6101 & 8443-12703 & 8991-9102 & 8943-3701 & 8568-1901 & 9485-1902 & 8322-3702 & 9486-6101 \\
9183-12704 & 8979-9101 & 8991-3702 & 8713-3702 & 8566-6101 & 8141-6104 & 8319-9102 & 7958-6104 \\
9182-3704 & 8313-3702 & 8990-12702 & 8604-12701 & 8555-6103 & 8483-6104 & 8319-6104 & 9486-6104 \\
9181-6103 & 8713-3702 & 8989-6103 & 8462-6102 & 8555-3704 & 8555-3702 & 8319-6103 & 9883-6103 \\
9181-3704 & 8948-6104 & 8989-12704 & 9045-12705 & 8555-12704 & 8485-3701 & 8317-6103 & 8948-6104 \\
9181-3702 & 8588-3701 & 8985-3703 & 8612-3704 & 8555-12701 & 9044-12703 & 8317-12701 & 8315-12702 \\
9181-12704 & 9486-6101 & 8984-3704 & 8253-3702 & 8554-6104 & 9505-12701 & 8315-6103 & 8713-3702 \\
9181-12703 & 9509-12705 & 8983-1902 & 8313-3704 & 8554-6103 & 9041-3701 & 8313-12705 & 8258-12704 \\
9181-12702 & 8547-12703 & 8983-12703 & 9881-6103 & 8554-6102 & 8948-6104 & 8309-12702 & 8483-6104 \\
9045-6103 & 9185-3703 & 8982-3701 & 8999-6103 & 8554-3702 & 8713-3702 & 8263-3702 & 8548-3701 \\
9045-6102 & 8131-6103 & 8980-12703 & 8244-6102 & 8553-6102 & 9487-12705 & 8262-9101 & 9034-12704 \\
9045-3704 & 8997-12704 & 8979-12701 & 8443-12703 & 8553-3703 & 9184-6102 & 8261-6101 & 8713-3702 \\
9045-3701 & 8713-3702 & 8978-9101 & 9881-12705 & 8553-12704 & 9034-1901 & 8261-3703 & 9182-1901 \\
9045-1902 & 8313-3702 & 8977-9101 & 8262-6103 & 8552-9102 & 8713-6103 & 8261-3702 & 8713-3702 \\
9045-12701 & 8943-3701 & 8977-3703 & 8713-3702 & 8552-9101 & 9038-12703 & 8259-3703 & 8718-6103 \\
9044-6104 & 9883-6104 & 8952-6102 & 8140-3702 & 8552-6103 & 8141-6104 & 8258-6102 & 9182-6103 \\
9044-3704 & 8258-3703 & 8952-3703 & 9881-3702 & 8551-3704 & 8440-3702 & 8257-3701 & 8600-6104 \\
9044-3703 & 8313-3704 & 8952-12702 & 9025-6103 & 8550-3704 & 8259-3702 & 8255-6104 & 8274-6103 \\
9044-3702 & 8455-6103 & 8952-12701 & 8600-3704 & 8550-12702 & 8978-12704 & 8255-6101 & 8567-6104 \\
9044-12705 & 8721-6102 & 8950-12705 & 8603-6102 & 8549-9101 & 8309-9101 & 8253-1901 & 8249-1902 \\
9044-12704 & 8332-6101 & 8948-6103 & 8980-3701 & 8549-12702 & 8602-12705 & 8249-6103 & 8938-6101 \\
9044-12702 & 8713-3702 & 8948-1902 & 9486-1902 & 8547-9101 & 8984-9101 & 8247-9102 & 8483-6104 \\
9044-12701 & 8313-6102 & 8947-6104 & 8938-6101 & 8486-3704 & 9870-3704 & 8247-6101 & 8551-1901 \\
9043-6103 & 9041-3701 & 8947-6101 & 8713-3702 & 8486-3701 & 8713-6103 & 8244-9102 & 9891-3701 \\
9043-3702 & 9486-1902 & 8947-3704 & 8326-9101 & 8485-9101 & 8326-3702 & 8244-6103 & 8483-6104 \\
9043-12703 & 8980-12702 & 8946-9102 & 8936-12702 & 8485-12703 & 8566-6104 & 8244-3704 & 9041-3701 \\
9043-12702 & 8261-3704 & 8946-6104 & 8948-6104 & 8483-6102 & 8313-3702 & 8244-3701 & 8993-3703 \\
9042-3701 & 8713-3702 & 8946-3703 & 8978-1901 & 8482-3703 & 8999-6103 & 8243-9102 & 8313-3702 \\
9041-6103 & 8948-6104 & 8946-1902 & 8313-3702 & 8482-1901 & 9485-1902 & 8150-6104 & 9028-12702 \\
9041-3704 & 8938-6104 & 8946-12703 & 8455-6103 & 8482-12702 & 9026-12704 & 8150-1901 & 9486-1902 \\
9039-6103 & 8713-3702 & 8946-12701 & 9487-12705 & 8481-9101 & 8481-6101 & 8149-12705 & 9487-12705 \\
9039-6101 & 8943-3701 & 8943-9101 & 9182-6101 & 8481-3704 & 8455-6103 & 8146-12705 & 9487-9101 \\
9039-1902 & 8613-3703 & 8943-3704 & 9041-3701 & 8466-3701 & 8718-3702 & 8146-12704 & 9184-9101 \\
9039-12701 & 9184-9101 & 8943-3703 & 8980-6102 & 8465-6101 & 8483-6104 & 8143-6104 & 9182-6103 \\
9038-12702 & 9034-12704 & 8943-3702 & 8140-3702 & 8465-12704 & 9872-12705 & 8143-6103 & 9038-6101 \\
9037-6104 & 8984-9101 & 8942-12702 & 8602-12705 & 8464-1902 & 7960-1902 & 8135-9101 & 8315-12702 \\
9037-6103 & 9487-1901 & 8942-12701 & 8313-3701 & 8462-3702 & 9041-3701 & 8135-6103 & 9487-12705 \\
9037-12704 & 8715-6104 & 8941-1901 & 8459-3703 & 8461-9101 & 8326-9101 & 8135-3703 & 8612-3704 \\
9036-3703 & 8139-9102 & 8938-9102 & 10001-9102 & 8461-3703 & 8713-3702 & 8135-12701 & 8459-3703 \\
9035-6103 & 9865-6102 & 8938-3704 & 9045-12705 & 8461-12701 & 8717-9102 & 8131-6102 & 8313-3702 \\
9035-3704 & 8980-3702 & 8937-1902 & 8987-1902 & 8459-6104 & 8551-6102 & 8131-12705 & 8330-12705 \\
9034-6104 & 9045-12705 & 8932-1902 & 8440-3701 & 8459-3701 & 8938-1901 & 8131-12702 & 8135-6102 \\
9033-9101 & 8274-6103 & 8725-6103 & 9182-6103 & 8456-6103 & 8939-6103 & 7992-12701 & 8309-9101 \\
9033-6104 & 8943-3701 & 8724-6101 & 8482-6103 & 8456-3702 & 9507-6101 & 7960-9102 & 9045-12705 \\
9033-6103 & 8326-9101 & 8724-12703 & 8249-6104 & 8456-3701 & 8253-3702 & 7960-1901 & 8547-1902 \\
9031-12703 & 8996-9102 & 8721-9102 & 8713-3702 & 8454-9102 & 8588-12701 & 7958-9102 & 8939-6103 \\
9029-9102 & 9038-6101 & 8721-6103 & 8984-12705 & 8454-6103 & 8274-6103 & 7958-3701 & 9002-6102 \\
9029-9101 & 8612-6101 & 8721-12703 & 8483-6104 & 8452-6102 & 9038-3704 & 7957-6103 & 8485-3703 \\
9029-12703 & 8936-12702 & 8721-12701 & 8312-9101 & 8452-3703 & 8948-6104 & 7957-12703 & 9042-6104 \\
9028-9102 & 8936-12705 & 8720-12702 & 8330-12705 & 8452-3702 & 8313-3702 & 7443-9102 & 9000-3703 \\
9028-3701 & 8315-12702 & 8717-6103 & 9002-6102 & 8447-6104 & 8948-9102 & 7443-6104 & 8548-3701 \\
9027-3704 & 8258-3703 & 8717-3702 & 8274-6103 & 8447-6102 & 8948-6104 & 10001-6104 & 8452-6103 \\
9026-6103 & 8484-6101 & 8717-1902 & 8555-3702 & 8447-3702 & 8713-3702 &  &  \\* \bottomrule

\caption{MaNGA plateifu for RDAGN host galaxies and their assigned control galaxy.}
\label{tab:RDAGN and controls}\\
\end{longtable}

\clearpage
\begin{longtable}[c]{@{}cccc@{}}
\toprule
AGN ID & \begin{tabular}[c]{@{}c@{}}RA\\ $[$deg$]$\end{tabular} & \begin{tabular}[c]{@{}c@{}}Dec\\ $[$deg$]$\end{tabular} & \begin{tabular}[c]{@{}c@{}}Justification for Excluding\\ from Final Sample\end{tabular} \\* \midrule
\endfirsthead
\multicolumn{4}{c}%
{{\bfseries Table \thetable\ continued from previous page}} \\
\endhead
\bottomrule
\endfoot
\endlastfoot
9031-9102 & 241.3982 & 44.20613 & Cube Quality: CRITICAL \\
9183-3701 & 119.968 & 38.24004 & Cube Quality: CRITICAL \\
8613-12705 & 255.6771 & 34.05999 & Cube Quality: CRITICAL \\
9182-9101 & 120.1768 & 40.0273 & Cube Quality: CRITICAL \\
8995-12704 & 175.5114 & 55.39062 & Cube Quality: CRITICAL \\
8995-3704 & 175.602 & 54.77419 & Cube Quality: CRITICAL \\
8995-12701 & 174.3928 & 54.85328 & Cube Quality: CRITICAL \\
8995-6104 & 176.508 & 55.41962 & Cube Quality: CRITICAL \\
8952-9102 & 205.6328 & 26.48724 & Cube Quality: CRITICAL \\
9024-6101 & 221.6665 & 33.30122 & Cube Quality: CRITICAL \\
8253-9101 & 157.6605 & 44.01272 & Cube Quality: BAD OMEGA \\
9486-9101 & 120.7992 & 39.88577 & Cube Quality: BAD OMEGA \\
8147-12705 & 117.9821 & 27.30297 & Cube Quality: BAD OMEGA \\
9035-9101 & 235.447 & 45.556 & Cube Quality: BAD OMEGA \\
8940-12704 & 122.0924 & 26.27565 & Cube Quality: BAD OMEGA \\
9181-12701 & 118.5709 & 38.22089 & Cube Quality: BAD OMEGA \\
8549-3703 & 241.4164 & 46.84656 & Cube Quality: BAD OMEGA \\
8329-12705 & 214.5477 & 44.47428 & Cube Quality: BAD OMEGA \\
8447-6103 & 206.173 & 40.4673 & Cube Quality: BAD OMEGA \\
9031-9101 & 239.1646 & 45.54078 & Cube Quality: BAD OMEGA \\
8439-12705 & 143.2881 & 49.05032 & Cube Quality: BAD OMEGA \\
8247-6103 & 136.72 & 41.40825 & Cube Quality: BAD OMEGA \\
8952-12704 & 205.2358 & 26.48672 & Cube Quality: BAD OMEGA \\
8568-12704 & 155.543 & 38.51782 & Cube Quality: BAD OMEGA \\
8612-12702 & 253.9464 & 39.31054 & Cube Quality: BAD OMEGA \\
8945-1902 & 174.4782 & 47.46635 & Cube Quality: BAD FLUX \\
8945-3704 & 175.1973 & 46.54049 & Cube Quality: BAD FLUX \\
8945-6102 & 173.7012 & 46.98995 & Cube Quality: BAD FLUX \\
8482-9101 & 241.7996 & 48.57256 & No radio emission at $>$ 3xrms \\
8603-6104 & 247.42 & 40.68695 & No radio emission at $>$ 3xrms \\
8552-6104 & 229.0521 & 45.23306 & No radio emission at $>$ 3xrms \\
8554-12701 & 182.2852 & 35.63581 & No radio emission at $>$ 3xrms \\
9029-12704 & 247.217 & 42.81201 & No radio emission at $>$ 3xrms \\
8326-3703 & 215.2749 & 48.30817 & No radio emission at $>$ 3xrms \\
8330-3702 & 203.8965 & 40.11109 & No radio emission at $>$ 3xrms \\
8257-1902 & 166.2978 & 46.10294 & No radio emission at $>$ 3xrms \\
8612-1902 & 254.0966 & 38.36347 & No radio emission at $>$ 3xrms \\
8459-12701 & 147.379 & 42.13029 & No radio emission at $>$ 3xrms \\
8712-1901 & 119.9737 & 55.37482 & No radio emission at $>$ 3xrms \\
8554-12703 & 182.7931 & 37.51535 & No radio emission at $>$ 3xrms \\
8326-6104 & 216.2561 & 47.95349 & No radio emission at $>$ 3xrms \\
8254-6103 & 162.9892 & 44.76013 & No radio emission at $>$ 3xrms \\
8595-3701 & 218.8973 & 50.18998 & No radio emission at $>$ 3xrms \\
9002-3701 & 222.8336 & 30.66383 & No radio emission at $>$ 3xrms \\
9028-3703 & 243.7375 & 30.75408 & No radio emission at $>$ 3xrms \\
8548-3703 & 243.044 & 47.90643 & No radio emission at $>$ 3xrms \\
8712-3704 & 122.2451 & 53.50988 & No radio emission at $>$ 3xrms \\
9865-12703 & 223.1398 & 50.92284 & No radio emission at $>$ 3xrms \\
9024-3702 & 221.792 & 33.21047 & No radio emission at $>$ 3xrms \\
8329-1901 & 214.4221 & 45.46582 & No radio emission at $>$ 3xrms \\
8253-6104 & 158.2514 & 42.92842 & No radio emission at $>$ 3xrms \\
8549-3704 & 243.1854 & 45.35201 & No radio emission at $>$ 3xrms \\
8721-12704 & 135.2365 & 54.95451 & No radio emission at $>$ 3xrms \\
8948-12702 & 164.9711 & 50.0152 & No radio emission at $>$ 3xrms \\
8993-3704 & 166.0866 & 46.0561 & No radio emission at $>$ 3xrms \\
8253-6102 & 158.533 & 42.80921 & No radio emission at $>$ 3xrms \\
8481-1902 & 237.6539 & 53.39062 & No radio emission at $>$ 3xrms \\
9869-9101 & 246.5913 & 40.91184 & No radio emission at $>$ 3xrms \\
9026-3704 & 251.3779 & 43.58164 & No radio emission at $>$ 3xrms \\
9883-12703 & 256.5416 & 33.60413 & No radio emission at $>$ 3xrms \\
8444-9101 & 200.6449 & 33.15709 & No radio emission at $>$ 3xrms \\
8253-6103 & 156.9885 & 43.31827 & No radio emission at $>$ 3xrms \\
7957-6102 & 258.2711 & 35.26862 & No radio emission at $>$ 3xrms \\
8252-12702 & 145.5308 & 48.15487 & No radio emission at $>$ 3xrms \\
8592-9102 & 224.4149 & 53.00634 & No radio emission at $>$ 3xrms \\
8601-3702 & 247.6121 & 40.72508 & No radio emission at $>$ 3xrms \\
8712-6104 & 121.5857 & 55.46234 & No radio emission at $>$ 3xrms \\
8716-3703 & 123.5062 & 52.75246 & No radio emission at $>$ 3xrms \\
8948-1901 & 165.7391 & 50.67024 & No radio emission at $>$ 3xrms \\
8480-9101 & 194.3831 & 28.47694 & No maps available from MaNGA \\
8479-12701 & 195.0339 & 27.977 & No maps available from MaNGA \\
8454-1902 & 154.7634 & 44.03303 & \begin{tabular}[c]{@{}c@{}}No control candidates with \textit{z} \\ or M$_{*}$ that varies $<$ 30$\%$\end{tabular} \\* \bottomrule
\caption{MaNGA galaxies excluded from the final RDAGN sample.}
\label{tab:excluded AGN}\\
\end{longtable}
\clearpage

\end{appendix}

\end{document}